\documentclass{statsoc}

\usepackage[a4paper]{geometry}
\usepackage{graphicx}
\usepackage{amsmath,amsfonts}
\usepackage{amssymb}
\usepackage{color}
\usepackage{natbib}
\usepackage{booktabs}
\usepackage{setspace}
\usepackage{verbatim, ulem}

\usepackage{float}
\floatstyle{plaintop}
\restylefloat{table}

\title[Predicting Phenotypes from Brain Connection Structure]{Predicting Phenotypes from Brain Connection Structure}
\author[Guha {\it et al.}]{Subharup Guha}
\address{Department of Biostatistics, University of Florida}
\author{Rex Jung}
\address{Department of Neurology, University of New Mexico Health Sciences Center}
\author{David Dunson}
\address{Department of Statistical Science, Duke University}
\author{\textcolor{white}{S.Guha, R.Jung, D.Dunson}}

\begin{document}
\begin{abstract}
This article focuses on the problem of predicting a response variable based on a network-valued predictor.  Our motivation is the development of interpretable and accurate predictive models for cognitive traits and neuro-psychiatric disorders based on an individual's brain connection network (connectome).  Current methods reduce the complex, high dimensional  brain network into low-dimensional pre-specified features prior to applying standard predictive algorithms.  These methods are sensitive to feature choice and inevitably discard important information. Instead, we  propose a nonparametric Bayes class of models that utilize  the entire adjacency matrix defining brain region connections  to adaptively detect predictive algorithms, while maintaining interpretability. The Bayesian Connectomics (BaCon) model class utilizes Poisson-Dirichlet processes to find a lower-dimensional, bidirectional (covariate, subject) pattern in the adjacency matrix.  The small $n$, large $p$ problem is transformed into a "small $n$, small $q$" problem, facilitating an effective stochastic search of the predictors. A spike-and-slab prior for the cluster predictors strikes a balance between regression model parsimony and flexibility, resulting in improved inferences and test case predictions. We describe basic properties of the BaCon model and develop efficient algorithms for posterior computation.  The resulting methods are found to outperform existing approaches and applied to a creative reasoning data set.
\end{abstract}

\section{Introduction}
Advances in non-invasive brain imaging technologies have made available brain  connectivity data  at increasingly greater  accuracies and spatial resolution. These advances have shifted the focus of neuroscience research  away from specialized brain regions   having independent effects on  cognitive functions \citep{Fuster_2000}   towards  brain connectivity networks (or \textit{connectomes}) in which cognitive processes operate as interconnected circuits \citep{Bressler_Menon_2010}. \cite{stirling_elliott_2008}, \cite{Craddock_etal_2013} and \cite{Wang_etal_2014} provide an overview of relevant technological developments, such as  diffusion tensor imaging (DTI), structural magnetic resonance imaging (sMRI) and magnetization-prepared gradient-echo (MP-RAGE).

This paper is  motivated by  investigations that seek to discover the relationship between brain connectivity structure and a subject-specific response, such as a quantitative creative reasoning score, the presence or absence of a neuropsychiatric disease, or type of ability. For individuals $i=1,\ldots,n$, the data consist of the categorical or quantitative response $y_i$ and the undirected  connectivity network among $V$ brain regions, represented by a binary $V \times V$ symmetric adjacency matrix, $\boldsymbol{A}_i=((a_{ij_1j_2}))$. For $j_1, j_2=1,\ldots,V$,   binary element $a_{ij_1j_2}$  equals 1 if and only if at least one white matter fiber connects brain regions $j_1$ and $j_2$ in subject $i$. In some investigations, a  vector of subject-specific covariates $\boldsymbol{r}_i$ is also available.

We focus on the  MRN-114  dataset available at {\tt \allowbreak http://openconnecto.me/data/ public/MR/}.
The responses $y_1,\ldots,y_n$ of   $n=114$ individuals are  creative reasoning scores measured using the composite
creativity index (CCI) \citep{Jung_etal_2010}. The brain region adjacency information for these individuals,  available from structural MP-RAGE and DTI brain scans \citep{Roncal_etal_2013}, consists of $V = 70$ network nodes corresponding to brain regions by the Desikan atlas \citep{Desikan_etal_2006} and equally divided between the left and right hemisphere.

The goal is to identify clusters of brain connections  operating in tandem, identify a sparse set of connections capable of explaining  individual variations in CCI,  and make reliable CCI predictions for out-of-the-bag individuals for whom only brain architecture information is available. These are challenging tasks, especially  because  the $70(70-1)/2=$ $2,415$    brain region pairs  overwhelm the number of~individuals, making it a ``small $n$, large $p$'' statistical problem.

\paragraph{Existing methods for  categorical responses in brain connectivity problems.}

Several methods have been developed for classification
based on the brain networks of individuals;
see \cite{Bullmore_Sporns_2009} and \cite{Stam_2014} for an overview. A majority of these methods reduce  individual connectivity information to prespecified summaries that characterize the network, e.g., number of connections, average path length, and clustering coefficient \citep{Rubinov_Sporns_2010}. These features are then  used in standard classification algorithms  such as support vector machines. Unfortunately, the results are  highly
sensitive to the chosen summary measures and  often ignore additional brain connectivity information   contributing to  individual or group differences. Refer to \cite{Arden_etal_2010} for examples of inconsistencies in analyses relating brain connectivity networks to creative reasoning.

An  alternative strategy avoids discarding useful connectome information by testing for differences between groups in each brain region pair, while  adjusting for multiple testing via false discovery rate (FDR) control \citep{Genovese_etal_2002}. However, because there are $V (V -1)/2$ distinct pairs of brain regions,  the number of tests is large when $V = 70$. Since they ignore network information,  these
univariate approaches tend to have low power \citep{Fornito_etal_2013} and substantially underestimate brain connectivity variation across groups. Some methods attempt to compensate for this by replacing the usual \cite{Benjamini_Hochberg_1995} approach with thresholding procedures
 utilizing  network information \citep[e.g.,][]{Zalesky_etal_2010}. Such
approaches require careful interpretation and their parameters must be meticulously chosen to give reliable results.

\cite{Durante_Dunson_2018} incorporate network information into their Bayesian model. This is accomplished by expressing the joint pmf of the data
$(y_i ,\boldsymbol{A}_i)$,  $i = 1,\ldots,n$, as the product of the marginal pmf of group  $y_i$ and the conditional pmf for   matrix $\boldsymbol{A}_i$ given the group. This approach facilitates testing of the association between  connectivity and the categorical
response, while borrowing information across subjects in learning the network structure.

Graph convolutional networks are  promising  approaches that  leverage the topology of brain networks. Recently,  \cite{liu2019auto}  developed a nonlinear latent factor model for summarizing the brain graph in
unsupervised and supervised settings. The  approach, called Graph AuTo-Encoding (GATE), relies on deep neural networks and is    extended  to regression with GATE (reGATE) to relate human phenotypes with brain structural connectivity.

\subsection{Inference goals}\label{S:goals}

This paper proposes a  nonparametric Bayes method capable of analyzing  categorical responses as well as quantitative responses such as continuous measurements and counts. For individual $i=1,\ldots,n$, the binary values $\{a_{ij_1j_2}: j_1>j_2 \text{ and } j_1, j_2=1,\ldots,V\}$ representing the pairwise connectivity of the brain regions are vectorized as  covariates $x_{i1},\ldots,x_{ip}$, where $p=V (V -1)/2$. This equivalent representation of the $n$ adjacency matrices gives an $n$ by $p$ matrix $\boldsymbol{X}$  consisting of $n$-variate column vectors denoted by $\boldsymbol{x}_j=(x_{1j},\ldots,x_{nj})'$,  $j=1,\ldots,p$.

From this perspective, the goals of the analysis  can be restated as follows: {\it{(i) Cluster detection:}} We wish to identify latent clusters of covariates having similar patterns for the subjects. As suggested by \cite{Bressler_Menon_2010}, these clusters may represent unknown cognitive
processes consisting of  brain region pairs  operating as interconnected circuits; {\it{(ii) Identification of sparse regression models:}} From the $p$ brain region pairs,
we wish to detect a reliable and parsimonious regression model for the responses;  {\it{(iii) Response prediction:}} Using the inferred regression model, we wish to  predict the responses of  additional subjects for whom only connectome information is available.
Because we are  interested in the relationship between the covariates and  responses, as a pre-processing step, we discard  any constant covariates (i.e., vectors of all $n$ zeros or all $n$ ones). In the  MRN-114 dataset, this gives  $p=1,374$ covariate vectors.

\paragraph{Some existing Bayesian approaches.} Outside the realm of connectome applications,
there are  general  Bayesian strategies for achieving one or more of the  analytical goals. However, since most of these techniques were not specifically designed for small $n$, large $p$ problems,    methods are being continually  developed to meet the statistical and computational challenges posed by newer applications and larger datasets.

 Bayesian clustering techniques typically rely on the ubiquitous Dirichlet process \citep[e.g. see][chap.\ 4]{muller2013bayesian}. 
 \cite*{Lijoi_Mena_Prunster_2007b} recommended Gibbs-type priors \citep*{Gnedin_Pitman_2005, Lijoi_Mena_Prunster_2007a}, such as Poisson-Dirichlet processes, for more flexibly fitting   cluster structures and  demonstrated their  utility  in some biomedical applications. More recently, \cite{guha2016nonparametric} introduced a clustering and variable selection technique for high dimensional  datasets with continuous covariates such as gene expression.  This technique is not directly applicable to structural connectivity datasets with binary   covariates, which
  require a very different approach. Motivated by these challenges, we propose a novel Bayesian clustering technique for binary covariates in small $n$, large $p$ problems that discovers the complex relationships between brain connectivity  and subject-specific phenotypes.

 \cite{OHara_Sillanpaa_2009} have reviewed Bayesian variable selection techniques  in linear and non-linear regression models.  For Gaussian responses,  common linear methods include stochastic search variable selection \citep{George_McCulloch_1993}, selection-based priors \citep{Kuo_Mallick_1997}, and shrinkage-based methods \citep{Park_Casella_2008, xu2015bayesian, griffin2010inference}. Empirical Bayes methods include \cite{yengo2014variable}, who model the  regression coefficients using a Gaussian mixture model. These regression methods make {strong} parametric assumptions   and do not  account for    collinearity commonly observed in high dimensional  datasets.
Some linear regression approaches allow nonparametric distributions for the error residuals \citep{hanson2002modeling,kundu2014bayes} and  regression coefficients \citep{bush1996semiparametric, maclehose2010bayesian}.

\begin{figure}
\begin{center}
\includegraphics[scale=0.4]{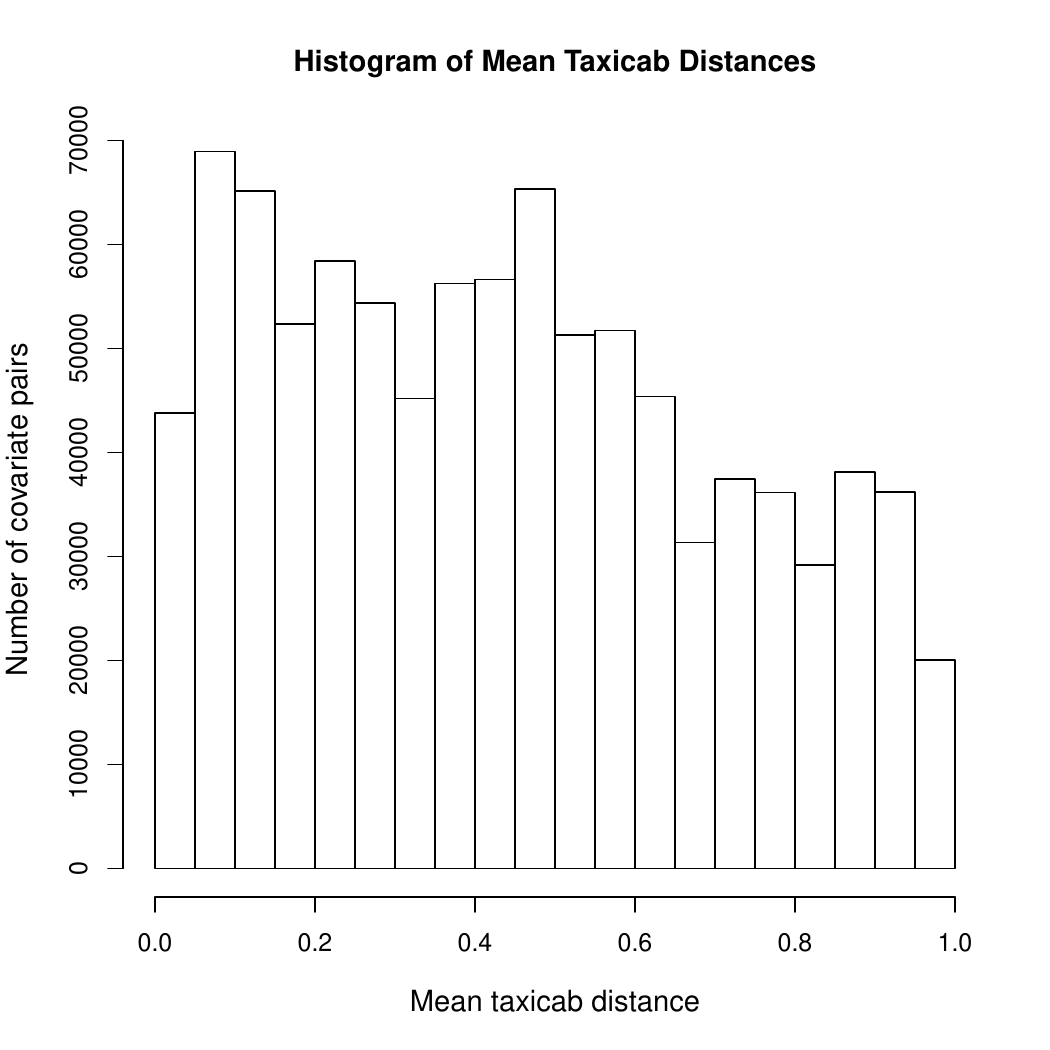}
\caption{For the MRN-114 dataset, mean taxicab distances between the $p=1,374$ non-constant covariate vectors of length $n=114$ each.}
\label{F:raw_taxicab}
\end{center}
\end{figure}

\paragraph{Challenges in high dimensional  settings.}
Variable selection is particularly challenging in  structural connectivity datasets because of the high degree of similarity  among the $p$ covariates.  Figure \ref{F:raw_taxicab}  displays the histogram of mean taxicab distances for the $p(p-1)/2$ $=943,251$  covariate pairs of the MRN-114 dataset. For binary-valued covariate vectors, a natural measure of similarity is the mean taxicab distance, which is a proportion lying between 0 and 1. A mean taxicab distance of 0 (1) corresponds to a perfect match (mismatch) between the $n$ elements of  two binary vectors.  The $25$th percentile of the mean taxicab distances in Figure \ref{F:raw_taxicab}  is 0.2018, and the distribution is skewed left, indicating substantial similarity between the covariate~vectors.

 This is a pervasive problem  not only in connectome datasets, but  more generally in small $n$, large $p$ problems. It occurs because the $n$-dimensional space of the covariate columns is saturated with the much larger number of covariates. Moreover, collinearity makes it  difficult to find a good set of predictors in regression settings.
 Collinearity also causes unstable  inferences  and  erroneous test case predictions \citep{Weisberg_1985}, rendering many of the aforementioned  techniques ineffectual in  brain connectivity~applications.

 This paper proposes BaCon (an acronym for \underline{Ba}yesian \underline{Con}nectomics), a fundamentally different approach  for connectome applications. The  technique specifies a  joint model for the covariates and responses and introduces new Bayesian nonparametric methodology for the unsupervised clustering of binary covariates. This innovation has the twin benefits of achieving dimension reduction and overcoming collinearity~issues.

\paragraph{Bidirectional clustering with regression variable selection and prediction.}

BaCon uses Poisson-Dirichlet processes (PDPs) to  group the $p$   columns of the covariate matrix  into $q$  latent clusters, where $q$ is much smaller than $p$. Each cluster consists of covariate columns that are  similar but not necessarily identical. The covariates belonging to a cluster are modeled as contaminated cluster-specific latent vectors;  the notion of ``contamination'' is  precisely defined in Section \ref{S:model}.
The taxicab distances between  covariates  belonging to a cluster are typically small, with occasional mismatches for a  small number of individuals.   The data are permitted to choose between PDPs and their special case, a Dirichlet process, for an appropriate covariate-to-cluster  allocation scheme.
 To flexibly capture the shared latent binary pattern  of the covariates within a cluster, each cluster allows the individuals to group differently  via nested Bernoulli mixtures. This  feature of the model is motivated by biomedical studies \citep[e.g.,][]{Jiang_Tang_Zhang_2004} which have broadly demonstrated that subjects tend to group differently under different biological processes.

 The proposed analytical framework detects a  lower-dimensional, bidirectional (covariate, subject)  clustering pattern in the binary covariates.
The small $n$, large $p$  problem is thereby transformed into a ``small $n$, small $q$'' problem,  facilitating an effective stochastic  search of the predictors. A spike-and-slab prior for the cluster predictors strikes a balance between regression model parsimony and~flexibility, resulting in improved
inferences and test case~predictions.

\begin{figure}
\hspace{-2 cm}
\includegraphics[scale=0.75]{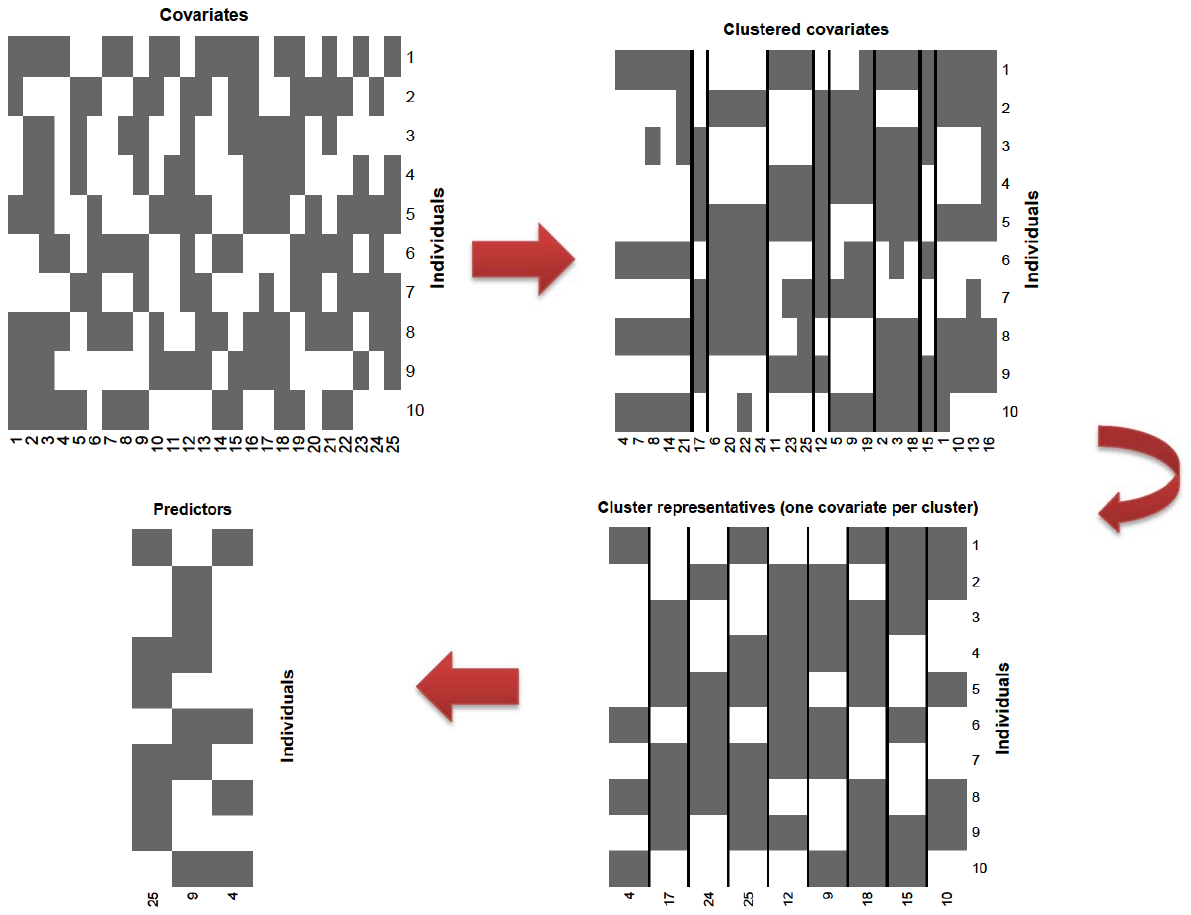}
\vspace{-2 cm}
\caption{Stylized example illustrating the key ideas of BaCon for $n=10$ subjects and $p=25$ covariates. The covariates belong to $q=9$ number of latent PDP clusters. The covariate indices are the column labels and the subjects are the row labels. Zero covariate values are shown in white and ones are shown in grey. The inferred regression relationship in the above situation is $Y=\beta_0+\beta_1X_{4}+\beta_6X_{9}+\beta_4X_{25}+\epsilon$, where the regression coefficient subscripts are the cluster labels of the predictors. See the text for further explanation.}
\label{F:toy_CoBaNa}
\end{figure}

Figure \ref{F:toy_CoBaNa}  illustrates  the main concepts using a toy example with $n=10$ subjects and $p=25$ covariates, with the zero covariates depicted as white and the ones as grey. The responses are continuous measurements, like the CCIs in the MRN-114 dataset. The plot in the upper left panel depicts the covariates. The posterior analysis averages over realizations of two basic stochastic steps:

 \begin{enumerate}
 \item\textbf{Clustering} \quad  The column vectors are assigned, based on  similarity, to $q=9$  PDP-Bernoulli mixture clusters. The shuffled covariate columns are plotted in the upper right panel. Notice that  covariates mapped to a  cluster are similar but not necessarily  identical.

     \item\textbf{Variable selection and regression} \quad One covariate called the \textit{cluster representative} is stochastically selected from each cluster. The regression predictors are chosen from this set.  The middle  panel displays the cluster representatives,  $\boldsymbol{x}_{4},\boldsymbol{x}_{17},\boldsymbol{x}_{24},\boldsymbol{x}_{25},\boldsymbol{x}_{12},$
$\boldsymbol{x}_{9},\boldsymbol{x}_{18},\boldsymbol{x}_{15}$, and $\boldsymbol{x}_{10}$. Only a few  representatives are response  predictors.
The predictors, $\boldsymbol{x}_{25},\boldsymbol{x}_{9}$, and $\boldsymbol{x}_{4}$, are shown in the lower panel. For  a zero-mean Gaussian error $\epsilon$, the regression equation  is $Y=\beta_0+\beta_1X_{4}+\beta_6X_{9}+\beta_4X_{25}+\epsilon$. The $\beta$ parameter subscripts are the cluster labels, e.g., coefficient $\beta_1$ is the effect of the first PDP cluster to which representative $\boldsymbol{x}_{4}$ belongs.
\end{enumerate}
In applications where an interpretable regression model is not of primary interest,  alternative variable selection strategies  discussed in Section \ref{S:predictors} can be applied.

The rest of the paper is organized as follows. Section~\ref{S:model} formally describes the BaCon model. Section \ref{S:MCMC} outlines the inference procedure. The substantial benefits and accuracy of BaCon are demonstrated by simulation studies in Section \ref{S:simulations}. The motivating connectome dataset, MRN-114, is analyzed in Section \ref{S:example}.


%

\section{The BaCon Model}\label{S:model}

The statistical model is motivated by the three-pronged goals of the analysis described in Section \ref{S:goals}. Dimension reduction in the $p=V(V-1)/2$ number of brain region pairs is achieved by Poisson-Dirichlet processes (PDPs), which allow  a  greater variety of clustering patterns than Dirichlet processes. The PDP allocations group the $p$ covariates into a smaller number of latent clusters. All covariate columns assigned to a cluster share a common $n$-variate pattern called the \textit{latent vector}. Occasionally, a random  misclassification may  occur at a given position of a covariate vector, causing the binary digit for that position (i.e., individual) to toggle relative to the latent vector element. From this perspective, the covariates may be regarded as contaminated versions of their cluster's latent vector.

\subsection{Covariate clusters}\label{S:PDP}

We assume that
each column vector $\boldsymbol{x}_j$ belongs to exactly one of $q \ll p$ latent  clusters, where the cluster memberships and $q$ are unknown.
For the covariate  $j=1,\ldots,p$ and cluster $k=1,\ldots,q$, the covariate-to-cluster assignment is determined by an \textit{allocation variable}
$c_j$, which equals $k$ if the $j$th covariate belongs to the $k$th latent cluster. The $q$ clusters are associated with  latent vectors $\boldsymbol{v}_1,\ldots,\boldsymbol{v}_q$ of length $n$, where each latent vector element $v_{ik} \in \{0,1\}$.

We model the covariate allocations as  partitions induced by
the \textit{two-parameter Poisson-Dirichlet process}, $\mathbb{PDP}\bigl(M, d \bigr)$, with  discount parameter $0 \le d < 1$ and mass parameter $M>0$. PDPs were introduced by \cite{Perman_etal_1992} and further studied by \cite{Pitman_1995} and \cite{Pitman_Yor_1997}. 
The PDP covariate-cluster assignment  can be described by an extension of the well-known Chinese restaurant process metaphor: Imagine customers, representing the covariates,  arriving at a restaurant. The first customer sits at a table labeled  1 (without loss of generality) so that $c_1=1$. The second customer may sit at table 1 with probability $(1-d)/(M+1)$ or sit at a new table having label 2 with probability $(M+d)/(M+1)$. The table eventually selected by customer 2 is recorded as $c_2$. The process proceeds  in this manner.
For customers $j=3,\ldots,p$, suppose there exist
   $q^{(j-1)}$ occupied tables among  the current customer-table assignments $c_1,\ldots,c_{j-1}$, with $n_{k}^{(j-1)}$ customers currently seated at the $k$th occupied table. The   probability that the $j$th customer sits at the $k$th table is~
\begin{align*}
P\left(c_j = k \mid c_1, \ldots, c_{j-1} \right) \propto
    \begin{cases}
        n_{k}^{(j-1)} - d 
         \quad &\text{if $k = 1,\ldots,q^{(j-1)}$}\\
        M + q^{(j-1)} \cdot d \quad &\text{if $k = q^{(j-1)} + 1$}\\
        \end{cases}
\end{align*}
where the event $c_j=q^{(j-1)} + 1$ in the second line corresponds to the $j$th customer selecting a new table, which is then assigned the label $q^{(j-1)} + 1$. Eventually, the sequence of customer-table selections is represented by  allocation variables $c_1,\ldots,c_{p}$. The  tables  occupied by the $p$ customers  represent the $q=q^{(p)}$ latent clusters, and the  customers seated at the $k$th occupied table represent the covariates allocated to the $k$th cluster.

Despite the sequential description of the extended Chinese restaurant metaphor, 
it can be shown that the $p$ allocation variables are  apriori exchangeable for PDPs, and more generally,    also for product partition models \citep{Barry_Hartigan_1993, Quintana_Iglesias_2003} and species sampling models \citep{Ishwaran_James_2003}.
The number of distinct clusters, $q$, is stochastically increasing in the PDP parameters $M$ and $d$. For  fixed $d$, all $p$ covariates are  assigned to singleton clusters (i.e., $q=p$) in the limit as $M \to \infty$.
 When $d = 0$, we obtain the  Dirichlet process with mass parameter $M$. Refer to \cite{Lijoi_Prunster_2010} for a detailed discussion of  Bayesian nonparametric~models, including Dirichlet processes and~PDPs.

PDPs allow  effective dimension reduction  in high dimensional  settings; the random number of clusters, $q=q^{(p)}$, is asymptotically equivalent to
\begin{align}
    \begin{cases}
        M  \log p        \qquad &\text{if $d = 0$} \quad \text{\it (Dirichlet process)}\\
        S_{d, M} \, p^d\qquad &\text{if $0 < d < 1$}\\
        \end{cases}\label{q}
\end{align}
where $S_{d, M}$ is a positive random variable. The number of clusters is asymptotically  smaller order than $p$,  resulting in  dimension reduction when $p$ is large. Equation~(\ref{q}) implies that the number of clusters for a Dirichlet process is smaller order than for a PDP. Dirichlet processes have been previously utilized
for dimension reduction; for example, see \cite{Medvedovic_etal_2004}, \cite{Kim_etal_2006}, \cite{Dunson_etal_2008}, and \cite{Dunson_Park_2008}.
The discount parameter $d$ is given  mixture prior $\frac{1}{2}\delta_0 + \frac{1}{2} U(0,1)$, where $\delta_0$ denotes a point mass at 0. 
Although we suspect that  connectome datasets may be more appropriately modeled with PDPs, this specification is appealing in allowing the model to adaptively simplify to a Dirichlet process when appropriate.

\vspace{-1cm}

\paragraph{Latent vector elements.}\quad The PDP prior specification is completed by   a \textit{base  distribution} in $\{0,1\}^n$ for each binary latent vector. We assume that the $nq$ elements  of  latent vectors  $\boldsymbol{v}_1,\ldots,\boldsymbol{v}_q$ are distributed as
\begin{equation}
v_{ik} \stackrel{iid}\sim \text{Bernoulli}(p_*), \qquad i=1,\ldots,n, \text{ } k=1,\ldots,q, \label{v}
\end{equation}
 allowing the clusters and individuals to communicate through a shared parameter which is given a conjugate prior:\
\begin{align}
p_* &\sim \text{Beta}\left(\lambda/2,\lambda/2\right), \quad \lambda>0. \label{G}
\end{align}
 The PDP base distribution is the $n$-fold product measure of this Bernoulli distribution. We denote  by  $q_*=1-p_*$ the prior probability of a latent vector element being 0.

The PDP allocations and mixture assumptions (\ref{v}) and (\ref{G}) for the latent vectors induce a nested  clustering of the $np$ covariates.
Unlike the clustering approaches for continuous covariates proposed by \cite{Fraley_Raftery_2002}, \cite{Quintana_2006} and  \cite{Freudenberg_etal_2010}, we do not assume that it is possible to \textit{globally} reshuffle the rows and columns of the data matrix to reveal a clustering pattern. Instead, somewhat similar to the nonparametric Bayesian local clustering (NoB-LoC) approach of \cite{Lee_Muller_Ji_2013}, we   cluster  the covariates locally using two sets of mixture models \citep{Hartigan_1990,Barry_Hartigan_1993,Crowley_1997}. However, there are significant differences in that our approach is primarily suited for binary rather than continuous covariates.  Furthermore, NoB-LoC relies solely   on two sets of Dirichlet processes,  whereas BaCon relies on Bernoulli mixtures nested within a~PDP.

\paragraph{Relating the covariates to the latent clusters}
Let the $j$th covariate be allocated to the $k$th cluster, so that $c_j=k$. As mentioned, the individual elements of column vector $\boldsymbol{x}_j$ arise as possibly corrupted versions of the $k$th latent vector's elements with a high probability of non-contamination (i.e.,  $x_{ij}=v_{ik}$). This results in similar patterns of  covariates   belonging to a cluster. Conditional on  latent vector element $v_{ik}=s$ $\in\{0,1\}$,  covariate $x_{ij}$ has the distribution
\begin{align}
P\left(x_{ij}=t \mid c_j = k, v_{i  k}=s, \boldsymbol{Q} \right)  =
    q_{s t}, \quad\text{where $t=0,1,$} \label{X}
\end{align}
for a $2 \times 2$ matrix of \textit{contamination probabilities} $\boldsymbol{Q}=((q_{st}))$.
High levels of agreement between the covariates and latent vectors are ensured by the diagonal elements of the matrix $\boldsymbol{Q}$ being close to 1. This, in turn, implies tight clusters with high levels of concordance  between   member covariates. 
From a broader statistical perspective, the idea of modeling the units in a cluster as contaminated versions of latent cluster-specific  characteristics is not new; for example, see \cite{dunson2009nonparametric} for a nonparametric Bayesian technique that allows dependent local clustering and borrowing of information using Dirichlet process~priors.

Row vectors $\boldsymbol{q}_0$ and $\boldsymbol{q}_1$ of the matrix $\boldsymbol{Q}$ sum to 1. They are assigned independent priors on the unit simplex in $\mathcal{R}^2$ as follows. Let $\mathcal{I}(\cdot)$ be the indicator function and let $\boldsymbol{1}_s$ be the $(s+1)$th unit vector in $\mathcal{R}^2$, i.e.,  with  the  $(s+1)$th element equal to 1 and the other elements equal to zero. For $s=0,1$,  row vector $\boldsymbol{q}_s$ has the expression
\begin{align}
 \boldsymbol{q}_s = (q_{s0},q_{s1}) &= r_s \boldsymbol{1}_s + (1-r_s) \boldsymbol{q}_s^*, \quad\text{where  row vector} \label{Q_rows}\\
  \boldsymbol{q}_s^* = (q_{s0}^*,q_{s1}^*)  &\sim \mathcal{D}_2\left(\alpha/2,\alpha/2\right), \quad\text{and}  \notag\\
   r_s &\sim \text{beta}(r_\alpha, r_\beta)\cdot\mathcal{I}(r_{s}>r^*),  \notag
 \end{align}
for prespecified constants $r^*$, $r_\alpha$ and $r_\beta$, and with $\mathcal{D}_2$ representing a Dirichlet distribution on the unit simplex in~$\mathcal{R}^2$.  Specification (\ref{Q_rows}), along with the assumption that $r^*>0.5$, guarantees that matrix $\boldsymbol{Q}$ is diagonally dominant. We refer to $r_s$ as the $s$th \textit{concordance parameter}. Since the concordance parameters determine the cluster separation, we  set $r^*=0.85$ to facilitate the detection of  tight  clusters.

\subsection{Regression and prediction}\label{S:predictors}

\paragraph{Continuous, categorical or count outcomes.} If the subject-specific responses are non-Gaussian, denote them by $w_1,\ldots,w_n$.
The Laplace approximation  \citep{Harville_1977} transforms the responses $w_i$ to independent regression outcomes $y_i$   having possibly approximate distributions, $N\left( \eta_i,\, \sigma_i^2\right)$. For an appropriate link function $g(\cdot)$, the normal mean $\eta_{i}=g(E[w_i])$. Laplace-type approximations are routinely used in exponential family models \citep{Zeger_Karim_1991,Albert_Chib_1993}. 
 Gaussian, Poisson, negative binomial,  and binomial responses all belong to this setting.
The approximation is exact for Gaussian responses (e.g., CCI responses in the MRN-114 dataset), which correspond to the identity link function and have a common  $\sigma=\sigma_i$ for all $n$ individuals.

\paragraph{Cluster-based covariate selection.}
Suppose $n_{k}$ covariates are allocated to the  $k$th cluster. 
To mitigate  collinearity effects, we assume that each cluster elects from its   member covariates a  \textit{representative}, denoted by $\boldsymbol{u}_k$. A subset of the $q$ cluster representatives, rather than of  the $p$ covariates, feature in an additive regression model.  The cluster representatives may be chosen in several different ways depending on the  application. Some possible options are:
\begin{enumerate}
\item[\textit{(a)}] Select with apriori equal probability one of the $n_k$  covariates belonging to the $k$th cluster. If covariate $s_k$ is selected as the  representative, then $c_{s_k}=k$ and $\boldsymbol{u}_k=\boldsymbol{x}_{s_k}$.

 \item[\textit{(b)}] We may find that some  covariates belonging to a cluster  closely resemble  the shared cluster pattern while others are barely in the cluster.  It may then be preferable to pick as the cluster representative the \textit{within-cluster median covariate}, the covariate having the minimal sum of distances to the other covariates.

\item[\textit{(c)}]     Select cluster-specific latent vector $\boldsymbol{v}_k$  as the cluster representative.
\end{enumerate}
Option \textit{(a)} is more relevant when practitioners are  interested in  interpretable models  identifying the effects of relevant regressors, i.e., brain region pairs. Option \textit{(b)} may be  preferred when the emphasis is more on identifying clusters of variables (e.g., cognitive processes)    jointly influencing the responses.

Extensions of  spike-and-slab priors \citep{George_McCulloch_1993, Kuo_Mallick_1997, Brown_etal_1998} are applied in selecting the regression predictors from  the $q$ cluster representatives:
\begin{align}
y_i &\stackrel{indep}\sim N\left( \eta_i,\, \sigma_i^2\right), \quad\text{where} \notag\\
\eta_i &= \beta_0 + \sum_{k=1}^q \gamma_{k}  \beta_k u_{ik}  \label{eta_i}
\end{align}

When the Laplace approximation is applied to the response $w_i$ to model regression outcome $y_i$,  variance  $\sigma_i^2$ may depend on~$i$, as in Poisson and binomial responses. If an additional  vector of known predictors $\boldsymbol{r}_i$ is available, it could be included in regression equation~(\ref{eta_i}) along with its regression coefficients. %

The linear predictor $\eta_i$ in expression (\ref{eta_i})  relies on a vector of cluster-specific indicators, $\boldsymbol{\gamma}=(\gamma_1,\ldots,\gamma_q)$. If $\gamma_{k}=0$, none of the covariates belonging to cluster~$k$ are associated with the response. If $\gamma_{k}=1$,  cluster representative $\boldsymbol{u}_k$ appears as a regressor  in equation (\ref{eta_i}).
The number of clusters associated with the response is then $q_1 =\sum_{j=1}^q \gamma_j$. The remaining  $q_0 =q-q_1$ clusters are not associated with the response.
For example, consider again Figure \ref{F:toy_CoBaNa}, where  one covariate from each cluster is the~representative, as described above in Option \textit{(i)}. Of the $q=9$ cluster representatives, $q_1=3$ are predictors and the remaining $q_0=6$  are non-predictors.

The following truncated
 prior for  indicator vector $\boldsymbol{\gamma}$ ensures model sparsity:
\begin{align}
[\boldsymbol{\gamma}] &\propto (1-\omega_1)^{q-q_1}\omega_1^{q_1} \cdot \mathcal{I}\biggl(q_1 <   n-1\biggr), \quad\text{where}\notag\\
\omega_1 &\sim \text{beta}(1,1).
 \label{gamma}
\end{align}
 Conditional on the variances $\sigma_i^2$ in equation (\ref{eta_i}), we assume a weighted g~prior for the regression coefficients of the predictors:
\begin{align}
\boldsymbol{\beta}_{\boldsymbol{\gamma}} | \boldsymbol{\Sigma} &\sim N_{q+1}\biggl(\boldsymbol{0}, \sigma_\beta^2({\boldsymbol{U}_{\boldsymbol{\gamma}}}'\boldsymbol{\Sigma}^{-1}\boldsymbol{U}_{\boldsymbol{\gamma}})^{-1}\biggr), \quad\text{where}\notag\\
\boldsymbol{\Sigma}&=\text{diag}(\sigma_1^2,\ldots,\sigma_n^2).
\label{Zellner}
\end{align}

\subsection{Justification of the clustering mechanism}\label{S:justifications}

We discuss the suitability of using PDPs as a  covariate clustering device and  an interesting consequence.

\begin{figure}
\begin{center}
\includegraphics[scale=0.4]{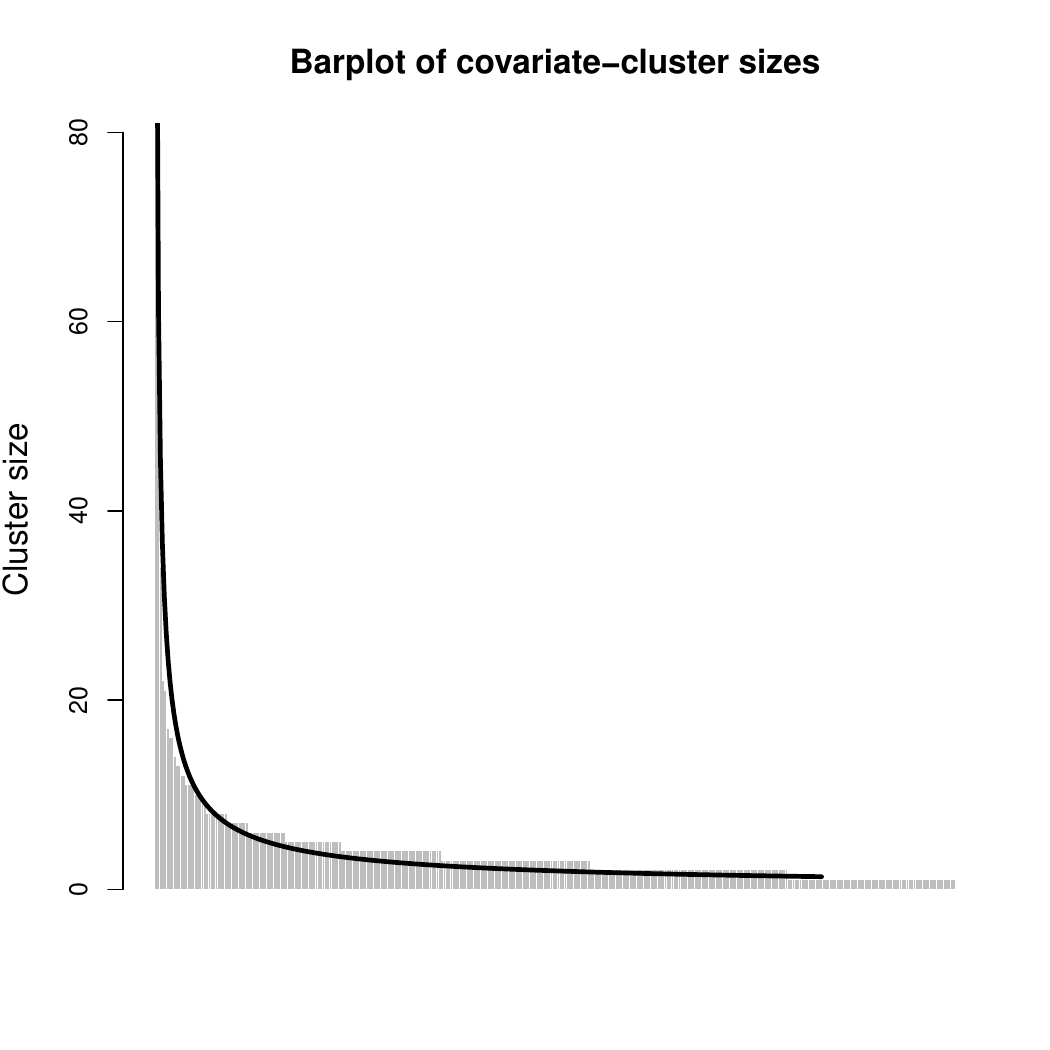}
\caption{Cluster sizes in the MRN-111 dataset detected by EDA. The best-fitting power law curve is overlaid in black.}
\label{F:eda_barchart}
\end{center}
\end{figure}

\paragraph{Empirical evidence against Dirichlet processes.} In an exploratory data analysis (EDA) of  brain region connectivity in the motivating MRN-114 dataset, the
 $p=1,374$ non-constant covariate vectors were grouped  in an ad hoc manner to detect the clusters. Specifically, we iteratively applied the k-means procedure to cluster the  covariates until the within-cluster median taxicab distances of the covariates were less than 0.4 for all the clusters.
The observed allocation pattern,  shown  in Figure \ref{F:eda_barchart},  is  highly uncharacteristic of Dirichlet processes; as is well known, Dirichlet processes are associated with relatively small numbers of clusters with exponentially decaying cluster sizes. The large number of clusters  ($\hat{q}=344$) and the predominance of  small clusters in Figure \ref{F:eda_barchart} suggest a non-Dirichlet covariate-cluster assignment.  PDPs are an attractive option because of their tractability, larger number of clusters, and the slower, power law decay of their cluster~sizes. For the MRN-114 dataset, the best-fitting power law function, $102.5k^{-0.74}$,  $k>1$, is shown in Figure \ref{F:eda_barchart}. 
 However, we prefer  a flexible general  specification that allows the data to choose between a PDP or Dirichlet process. Hence, as described in 
Section \ref{S:PDP}, we choose a mixture prior for the PDP parameter $d$ with a point mass at zero corresponding to the Dirichlet process.

\paragraph{Theoretical consequences and justifications for a PDP model.}

BaCon's nested mixture model cluster structure has some interesting consequences. The $n$-variate base distribution of the PDP is discrete, and there is a positive probability that two clusters have  identical latent vectors. However, an upper bound of the probability that two or more of the $q$ PDP clusters have identical latent vectors  is   ${q \choose 2} \bigl(p_*^2+q_*^2\bigr)^n$, with  $p_*$ and $q_*$  defined in expression (\ref{G}). Applying  asymptotic relationship (\ref{q}), we find that the upper bound approaches 0 as the dataset grows, provided the number of covariates, $p$, grows at a slower-than-exponential rate with $n$. Even for moderate-sized datasets with $n=50$   and $p=250$,  all the  latent vectors were distinct in our data analyses and simulation studies. Consequently, it is reasonable to assume that all BaCon clusters have unique features in structural connectivity~datasets.

\section{Posterior inference}\label{S:MCMC}
 Starting with ad hoc estimates, the BaCon model parameters are iteratively updated by MCMC methods. The post--burn-in MCMC sample is used for posterior inference. As a benefit of having a coherent stochastic model,  we are able to appropriately incorporate uncertainty into the inferences.
 Due to the computationally intensive MCMC procedure, the analysis is performed in separate steps, consisting of dimension reduction in the  covariates followed by variable~selection.

 \begin{enumerate}
 \item[\textbf{Step 1}] Focusing only on the binary connectivity information for  the brain regions:

      \begin{enumerate}
    \item[\textit{Step 1(i)}] The allocation variables,  latent vector elements, and all model parameters directly  related to the covariates are updated until the MCMC chain converges.  Section \ref{S:MCMC_c} describes  Gibbs sampling updates for the $p$  allocation variables. Section \ref{S:MCMC_v} specifies a Gibbs sampler for the latent vector elements.  Sections \ref{S:MCMC_Q} describes a Gibbs sampler for the contamination probability matrix, $\boldsymbol{Q}$. The remaining hyperparameters, such as the PDP discount parameter $d$, are generated using standard MCMC techniques.

Monte Carlo estimates  are  computed for the posterior probability of clustering for each pair of covariates. Following \cite{Dahl_2006}, these probabilities are  used to compute a point estimate, called the \textit{least-squares  allocation}, for the PDP assignments.

    \item[\textit{Step 1(ii)}] Conditional on the least-squares  allocation consisting of $\hat{q}$ PDP clusters, a second MCMC sample of the $n\hat{q}$ latent vector elements is generated. An estimate of these binary latent vector elements,  called the \textit{least-squares configuration}, is evaluated by again applying  the technique of \cite{Dahl_2006}.
  \end{enumerate}

 \item[\textbf{Step 2}] Finally, using the responses, and conditional on the least-squares allocation and  least-squares configuration, the regression predictors and any latent regression outcomes are generated to obtain a third MCMC sample. Response predictions are also made for test set individuals (if any).
 \end{enumerate}

\subsection{MCMC procedure}

 \subsubsection{Covariate-to-cluster allocation}\label{S:MCMC_c}

For the $j$th covariate column, we perform Gibbs sampling updates of  PDP allocation variable $c_j$, $j=1,\ldots,p$. The simulation strategy consists of the following steps:

 \begin{enumerate}

\item Discard  parameters exclusively related to the $j{th}$ covariate. Let
   $q^-$ be the number of clusters among the remaining $(p-1)$ allocation variables, with the $k$th cluster containing  $n_{k}^-$ number of covariates.
The $j$th covariate may join one of the existing $q^-$ clusters or  open a new cluster having the label  $(q^-+1)$. We evaluate the probabilities of these events and update parameter $c_j$ as described in Steps (\ref{existing_clust}) -- (\ref{generate_c_j}).

       \item\label{existing_clust} For each of the existing clusters, i.e., for $k=1,\ldots,q^-$, compute:

    \begin{enumerate}

    \item\label{psi_1} \underline{\textbf{Transition counts for the cluster-covariate combination}}\label{MCMC:psi} \quad 
        Compute  matrix $\boldsymbol{N}^{(jk)} =$ $((n_{st}^{(jk)}))$,  the $2 \times 2$  table of transition counts, defined as
    \begin{equation} n_{st}^{(jk)}= \sum_{i=1}^n\mathcal{I}(v_{i k}=s, x_{ij}=t),
    \qquad\text{for $s,t=0,1$}. \label{n_jk}
    \end{equation}

    \item \underline{\textbf{Posterior probability that allocation variable $c_j=k$}} \quad
        The posterior probability of the $j$th covariate belonging to the $k$th cluster is proportional to
        \begin{align}
       &\xi_{jk}  = (n_{k}^{-} - d) \cdot \prod_{s=0,1} \prod_{t=0,1} q_{st} ^{n_{st}^{(jk)}} \quad\text{for $k=1,\ldots,q^-$}. \label{prob.cj.1}
        \end{align}
    \end{enumerate}

   \item  \underline{\textbf{Posterior probability that allocation variable $c_j=q^-+1$}} \quad The posterior probability of the $j{th}$ covariate opening a new cluster is proportional to
         \begin{align}
       &\xi_{j\, (q^-+1)}  = (M + q^{-} d) \cdot \prod_{t=0,1}\bigl(q_* q_{0t}+p_* q_{1t}\bigr) ^{n_{t}^{(j(q^-+1))}} \label{prob.cj.2}
        \end{align}
        where $n_{t}^{(j(q^-+1))}=\sum_{i=1}^n\mathcal{I}(x_{ij}=t)$, and   $q_*$ and $p_*$ are defined in relation (\ref{G}).

    \item \underline{\textbf{Generation of allocation variable $c_j$}}\label{generate_c_j} \quad Using the values computed in expressions (\ref{prob.cj.1}) and (\ref{prob.cj.2}), evaluate the  constant $\xi_{j}$ that normalizes to probabilities the values $\xi_{j1},\ldots,\xi_{j\, (q^-+1)}$. That is, $\xi_j=1/\sum_{k=1}^{q^-+1}\xi_{jk}$. Set the allocation variable $c_j$ equal to $k$ with probability equal to $\xi_j\cdot \xi_{jk}$, or $k=1,\ldots,(q^-+1)$. 
        If $k=(q^-+1)$,  also generate the latent vector $\boldsymbol{v}_{q^-+1}$ for the new cluster: conditional on   $p_*$  and   matrix $\boldsymbol{Q}$,  the $n$ elements of  vector $\boldsymbol{v}_{q^-+1}$ have a posteriori independent Bernoulli distributions.

\end{enumerate}

 \subsubsection{Latent vector elements}\label{S:MCMC_v}

Among  allocation variables $c_1,\ldots,c_p$,  suppose there are $q$ clusters with
  cluster $k$ consisting of $n_{k}=\sum_{j=1}^p \mathcal{I}(c_j=k)$  covariates. The sufficient statistics for updating the latent vector elements is the $n$ by $ q$ matrix of counts, $\boldsymbol{W}=$ $((w_{ik}))$, where $w_{ik}=$ $\sum_{j:c_j=k}\mathcal{I}(x_{ij})$.
Conditional on  parameter $p_*$ and on the matrices $\boldsymbol{Q}$ and $\boldsymbol{W}$, the $nq$  latent vector elements have independent Bernoulli full conditional distributions.

\subsubsection{Gibbs sampler for contamination probability matrix $\boldsymbol{Q}$}\label{S:MCMC_Q}

Using the row vectors $\boldsymbol{q}^*_s=(q_{s0}^*,q_{s1}^*)$ and concordance parameters of relation (\ref{Q_rows}), let the matrix $\boldsymbol{Q}^*=((q_{st}^*))$ and concordance parameter vector, $\boldsymbol{r}=(r_0,r_1)'$. From relation~(\ref{Q_rows}), we find that updating  the  matrix $\boldsymbol{Q}$ is  equivalent to  a posteriori generating vector $\boldsymbol{r}$ followed by  updating  matrix $\boldsymbol{Q}^*$ conditional on   $\boldsymbol{r}$. The details are described below.

Comparing each cluster's latent vector to its allocated covariates, evaluate matrix $\boldsymbol{N} =$ $((n_{st}))$,  the $2 \times 2$ table of transition counts, which is the  sufficient statistic for updating matrix $\boldsymbol{Q}$. That is, the
  transition count
\[ n_{st}= \sum_{i=1}^n\sum_{j=1}^p\mathcal{I}(v_{i c_j}=s, x_{ij}=t),
\qquad\text{for $s,t=0,1$}.
 \]

\paragraph{Updating concordance parameter vector $\boldsymbol{r}$}\quad
For $s=0,1$, define the pmf
\begin{equation}
h_s(v)=
\begin{cases}
g_s(v)/\sum_{u=0}^{n_{ss}}g_s(u) \quad &\text{if $v = 0,\ldots,n_{ss}$},\label{h_s}\\
0 \quad&\text{otherwise},
\end{cases}
\end{equation}
which relies on  non-negative functions $g_0(\cdot)$ and $g_1(\cdot)$  having the definition:
\[
g_s(v)=
\begin{cases}
{n_{ss} \choose v} \frac{B(\boldsymbol{n}_s + \frac{\alpha}{2}\boldsymbol{1}-v\boldsymbol{1}_s)}{B(v+r_\alpha, N_s-v+r_\beta)}
\tilde{F}(r^* \mid v+r_\alpha, N_s-v+r_\beta)\quad &\text{if $v = 0,\ldots,n_{ss}$},\\
0 \quad&\text{otherwise},
\end{cases}
\]
where $\boldsymbol{n}_s$ denotes the $s$th row of matrix $\boldsymbol{N}$, $N_s=\sum_{t=0,1} n_{st}$ is the matrix's $s$th row sum, and $\boldsymbol{1}$ is the bivariate vector of ones. As defined  in equation (\ref{G}), $\boldsymbol{1}_s$ is the $(s+1)$th unit vector in $\mathcal{R}^2$. The survival function (i.e., 1 -- cdf) for the beta distribution with parameters $(v+r_\alpha)$ and $(N_s-v+r_\beta)$ is denoted by  $\tilde{F}(\cdot \mid v+r_\alpha, N_s-v+r_\beta)$. For a bivariate vector $\boldsymbol{a}=(a_1,a_2)$,  beta function $B(\boldsymbol{a})=$ $B(a_1,a_2)=$ $\prod_{s=0,1} \Gamma(a_{s+1})/\Gamma(\boldsymbol{a}'\boldsymbol{1})$.

Then, as shown in  the Appendix, the concordance parameters are a posteriori independently distributed as truncated beta distributions:
\begin{align}
r_s \mid \boldsymbol{X}, V_s, \cdots &\stackrel{indep}\sim \text{beta}(V_s+r_\alpha,N_s-V_s+r_\beta)\cdot \mathcal{I}(r_s > r^*), \quad\text{where} \notag\\
V_s &\stackrel{indep}\sim h_s(\cdot), \qquad s=0,1. \label{[r|X]}
\end{align}

\paragraph{Updating matrix $\boldsymbol{Q}^*$ conditional on concordance parameter vector $\boldsymbol{r}$}\quad For $s=0,1$, define the pmf
\begin{equation}
l_s(v)=
\begin{cases}
l_s^*(v)/\sum_{u=0}^{n_{ss}}l_s^*(u) \quad &\text{if $v = 0,\ldots,n_{ss}$},\label{l_s}\\
0 \quad&\text{otherwise},
\end{cases}
\end{equation}
where the non-normalized function
\[
l_s^*(v)=
\begin{cases}
{n_{ss} \choose v}  B(\boldsymbol{n}_s + \frac{\alpha}{2}\boldsymbol{1}-v\boldsymbol{1}_s)\rho_s^{v} \quad &\text{if $v = 0,\ldots,n_{ss}$},\\
0 \quad&\text{otherwise},
\end{cases}
\]
and this depends on the concordance parameter $r_s$ through $\rho_s=r_s/(1-r_s)$.
Then the row vectors $\boldsymbol{q}^*_s$ of matrix $\boldsymbol{Q}^*$ are a posteriori independently  distributed as
\begin{align}
\boldsymbol{q}^*_s \mid \boldsymbol{X}, r_s, U_s, \cdots &\stackrel{indep}\sim \mathcal{D}_2\left(\boldsymbol{n}_s + \frac{\alpha}{2}\boldsymbol{1}-U_s\boldsymbol{1}_s\right), \quad\text{where} \notag\\
U_s &\stackrel{indep}\sim l_s(\cdot), \qquad s=0,1. \label{[Q*|X,r]}
\end{align}
Refer to  the Appendix for the derivation.

\section{Simulation studies}\label{S:simulations}

\subsection{Cluster-related inferences}\label{S:simulation2}

As  discussed in Section \ref{S:justifications}, the  PDP allocations can be interpreted as clusters with unique characteristics. We investigate BaCon's accuracy as a clustering procedure using simulated covariates for which the true clustering pattern is known. 
In general, when allocating $p$ objects to an unknown number of clusters using mixture models, the non-identifiability and redundancy of the inferred clusters have been extensively studied \citep[e.g., see][]{Fruhwirth-Schnatter_2006}. Some partial solutions  are available within the Bayesian paradigm. For example, instead of assuming exchangeable  component parameters for finite mixture models, \cite{Petralia_etal_2012}  invent a repulsive process that  leads to a smaller number of better separated and more interpretable
clusters. \cite{Rousseau_Mengersen_2011} show that in  over-fitted finite mixture models, asymptotic emptying of the
redundant components  is achieved by a carefully chosen prior.

In brain connectome applications, the aforementioned  asymptotic results  assume that the number of rows of covariate matrix $\boldsymbol{X}$ remains fixed as the number of columns  tends to $\infty$. These results do not  guarantee that the BaCon model  correctly detects even the \textit{number} of covariate clusters. Nevertheless, the following simulation studies suggest a much stronger result: covariates that (do not) cluster under the true process also tend (not) to cluster a posteriori. The key intuition is that if $n$   also grows with $p$,  two  covariate column vectors that actually belong to different clusters are eventually   separated enough for the BaCon method to allocate them to different clusters. Similarly, the allocations of  covariates  belonging to the same cluster are  correctly called when  $n$ and $p$ are both large. This remarkable phenomenon has been documented in other high dimensional settings;   \cite{guha2016nonparametric} offer a formal explanation for continuous covariates such as gene expression datasets in cancer research.

\subsubsection{Data generated from the BaCon model}\label{S:simulation_PDP}
Binary covariates for $n=100$ individuals  and $p=250$ covariates were generated from the proposed model, and the inferred clusters were  compared with the  truth. The true parameters of the generating model were chosen to approximately match the estimates for the MRN-111 dataset. For each of 25 synthetic datasets,  with the true concordance parameters in relation~(\ref{Q_rows}), determining  cluster separation,   taking  the values  $r_0^{(0)}=$  $r_1^{(0)} \in$  $\{0.875, 0.925, 0.975\}$, the binary covariate matrix $\boldsymbol{X}$ was generated as  follows.:

\begin{enumerate}

 \item \textbf{True allocation variables:} We generated partitions $c_1^{(0)},\ldots,c_p^{(0)}$   induced by a PDP with  discount parameter $d^{(0)}=0.4$ and mass parameter $\alpha_1=20$. The true number of clusters, $Q_0$, was  computed for the partition.

\item \textbf{Latent vector elements:} For $i=1,\ldots,n$ and $k=1,\ldots,Q_0$, we simulated elements $v_{ik}^{(0)} \stackrel{iid}\sim \text{Bernoulli}(p^{(0)})$ with $p^{(0)}=5/7$.

\item \textbf{Contamination probability matrix:} As indicated in expression (\ref{Q_rows}), for $s=0,1$, we generated bivariate vector ${\boldsymbol{q}_s^*}^{(0)} \stackrel{iid}\sim \mathcal{D}_2\left(1,1\right)$. We computed the $s$th row vector of matrix~$\boldsymbol{Q}^{(0)}$  as $\boldsymbol{q}_s^{(0)} = $ $(q_{s0}^{(0)},q_{s1}^{(0)}) = r_s^{(0)} \boldsymbol{1}_s + (1-r_s^{(0)}) {\boldsymbol{q}_s^*}^{(0)}$. 

\item\label{X_step} \textbf{Binary covariates:} For individual $i=1,\ldots,n$ and covariate $j=1,\ldots,p$, let the true latent vector element be denoted by $g_{ij}$. That is, $g_{ij}=v_{ik}^{(0)}$ where $k = c_j^{(0)}$.  Each covariate was independently generated as
$x_{ij}  \sim \text{Bernoulli}(q_{g_{ij} 1}^{(0)})$.
\end{enumerate}

  There were no responses in this study.
  Each artificial dataset was analyzed using the BaCon methodology assuming all  parameters to be unknown.
   The accuracy of the inferred covariate-cluster allocation was evaluated by the \textit{proportion of correctly clustered covariate pairs},
        \[
        \tau = \frac{1}{{p \choose 2}}  \sum_{j_1 \neq j_2 \in \{1,\ldots,p\}} \mathcal{I}\biggl(\mathcal{I}({c}_{j_1}={c}_{j_2})=\mathcal{I}(c_{j_1}^{(0)}=c_{j_2}^{(0)})\biggr).
        \]
        This measure was estimated as an  MCMC empirical average, $\hat{\tau}$, with
        a high value  indicative of high  clustering accuracy.

       The second column of Table \ref{T:varkappa2} displays the percentage $\hat{\tau}$ for BaCon averaged over the 25 independent replications as cluster separation changes. The posterior inferences were found to be  robust to the contamination levels, i.e., concordance parameter. On average, less than 34.2  pairs  were incorrectly clustered out of the ${250 \choose 2}=$ 31,125 different covariate pairs, and so $\hat{\tau}$ was  greater than 99.89\%. Furthermore, in every  dataset, $\hat{q}$, the estimated number of clusters  of the least-squares  allocation was exactly equal to $Q_0$, the true number of PDP~clusters.

As a straightforward competitor to the BaCon technique, we  applied the k-means algorithm to  group the $p$ columns of  matrix $\boldsymbol{X}$ into the true number, $Q_0$, of PDP clusters. The percentage of correct allocations,  averaged over the 25 independent replications, are displayed in Column~3 of Table \ref{T:varkappa2}. Although setting the number of k-means clusters equal to $Q_0$ gives the k-means algorithm an unrealistic advantage, BaCon significantly outperformed it with respect to clustering accuracy.

We assessed
 BaCon's ability to discriminate between PDPs and Dirichlet processes using the log-Bayes factor, $
    \log\left(P[d>0|\boldsymbol{X}]/P[d=0|\boldsymbol{X}]\right)$. With  set $\Theta^*$ representing all model parameters except $d$, we obtain by Jensen's inequality that the log-Bayes factor in favor of PDP models exceeds $E\left(\log\left(\frac{P[d>0|\boldsymbol{X},\Theta^*]}{p[d=0|\boldsymbol{X},\Theta^*]} \right) \mid \boldsymbol{X} \right)$. Unlike the log-Bayes factor, this quantity can be easily  estimated using  only the post--burn-in MCMC sample. Furthermore, it  provides a lower bound for the log-Bayes factor itself, rather than a lower bound for the marginal log-likelihoods from which the log-Bayes factor is derived.
    The second column of  Table \ref{T:varkappa} displays averages and standard deviations of the log-Bayes factor's
     lower bound for the 25 datasets. These numbers correspond to 
Bayes factors in favor of PDP models significantly exceeding $e^{45}$, and are extreme evidence in favor of  PDP~allocations, i.e.,  the truth.

 Reliable posterior inferences were also achieved for the PDP discount parameter, $d \in [0,1)$. Column 3 of  Table \ref{T:varkappa} displays the 95\% posterior credible intervals for $d$. The posterior inferences are much more precise than the prior, with each CI containing the true value of~$d_0=0.4$. No posterior mass is assigned to Dirichlet process models in spite of the prior probability, $\mbox{Pr}(d=0)=0.5$.

\begin{table}
\caption{When the data were generated from the BaCon model,  the proportion of correctly clustered covariate pairs for different values of the true concordance parameter for the two competing methods. The standard errors are shown in  parentheses.} \label{T:varkappa2}
\vspace{.2 in}
\centering
\begin{tabular}{ c   | c |c  }
\hline\hline
 \textbf{Concordance} &\multicolumn{2}{c}{\textbf{Percent $\hat{\tau}$}} \\
  \cline{2-3}
 \textbf{parameter} &\textbf{BaCon}  &\textbf{K-Means}\\
\hline
0.875 &99.890 (0.011) &98.850  (0.165) \\
0.925 &99.896 (0.008) &98.973 (0.119) \\
0.975 &99.891 (0.010) &99.302 (0.116) \\
\hline\hline
\end{tabular}
\end{table}

\begin{table}
\caption{When the data were generated from the BaCon model,  column 2 presents the average lower bound of the log-Bayes factor of PDP models, relative to Dirichlet process models, for different values of the true concordance parameter. Standard deviations for the 25 independent replications are shown in  parentheses. Column 3 displays  95 \%  posterior credible intervals for the PDP discount parameter $d$. }
\label{T:varkappa}
\vspace{.2 in}
\centering
\begin{tabular}{ c   | c  | c}
\hline\hline
 \textbf{Concordance}  &\textbf{Lower  bound } &\textbf{95\% C.I.}\\
 \textbf{parameter}  &\textbf{of  log-BF} &\textbf{for $d$} \\
\hline
0.875  &73.232  (13.819) &(0.331, 0.527)\\
0.925  &73.327 (13.250) &(0.332, 0.536)\\
0.975 &80.100 (17.598) &(0.349, 0.541)\\
\hline\hline
\end{tabular}
\end{table}

\begin{figure}
\begin{center}
\includegraphics[scale=0.41]{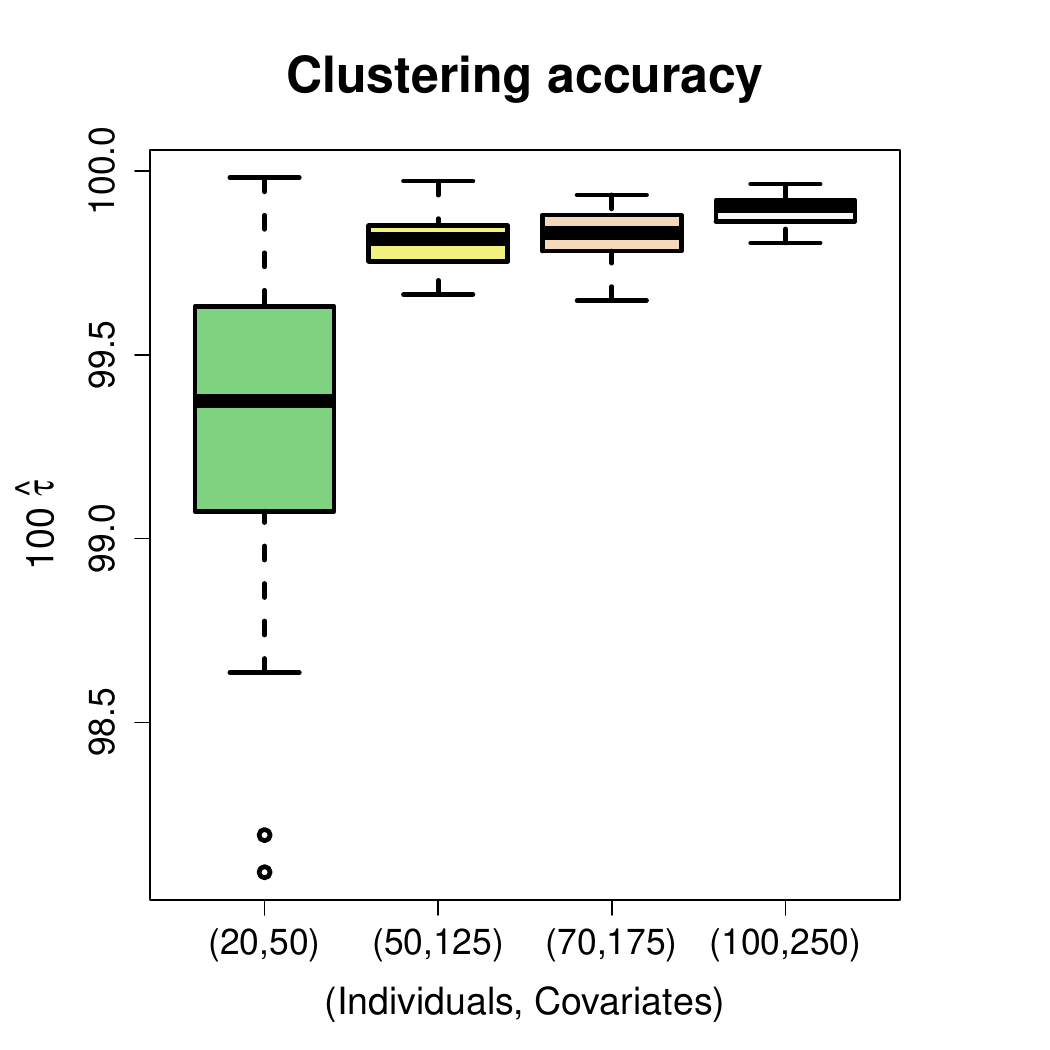}
\includegraphics[scale=0.41]{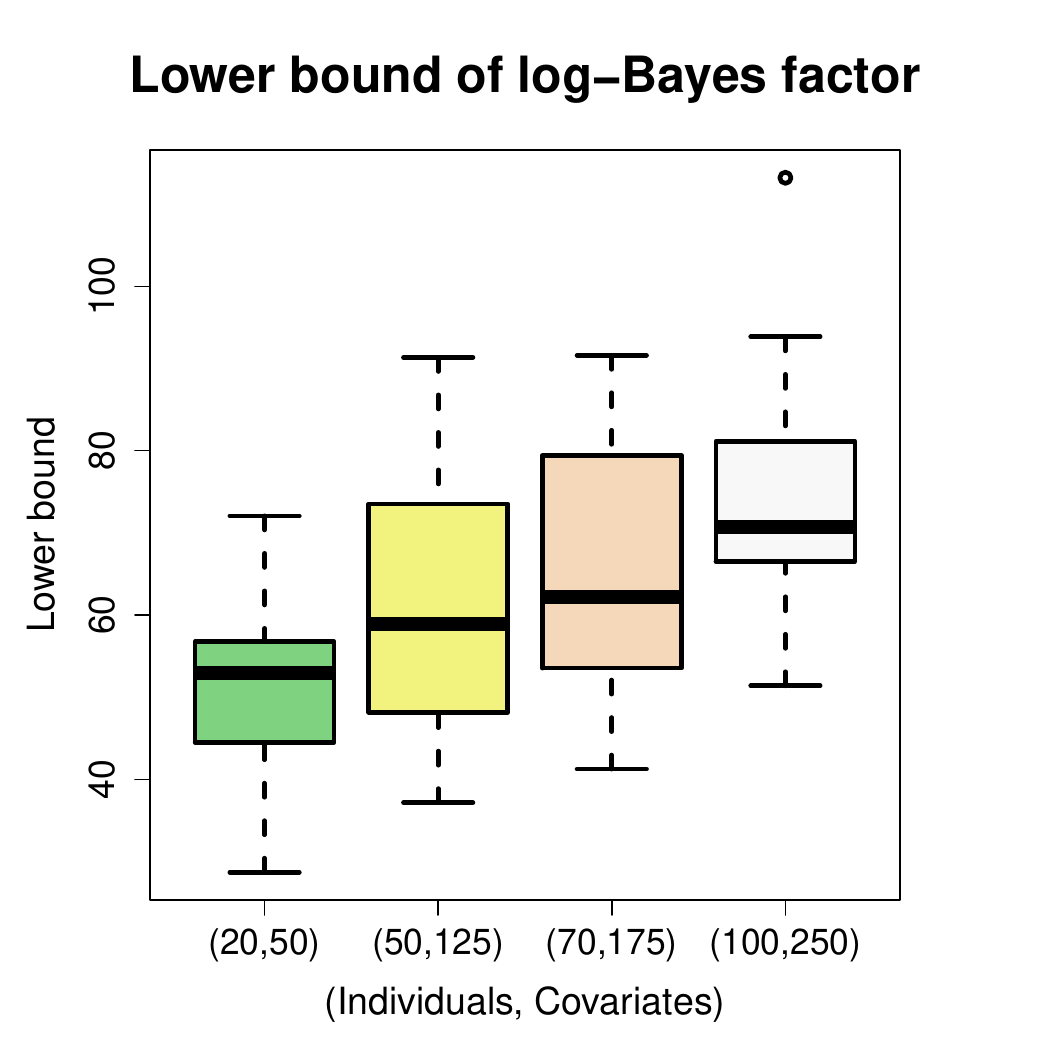}
\caption{Posterior inferences for artificial datasets of increasing dimension corresponding to true concordance parameters $r_0^{(0)}=r_1^{(0)}= 0.925$. See the discussion in Section \ref{S:simulation2}.}
\label{F:asymptotic}
\end{center}
\end{figure}

 \paragraph{Asymptotic convergence}\quad We compared the estimated and true values of  various features of the BaCon model as  the number of individuals  $n$ and covariates $p$ increased. Specifically, we implemented the  generation and inference strategy described above for the  increasing $(n,p)$ tuples,  $(20,50)$, $(50,125)$, $(70,175)$, and $(100,250)$. The results are summarized in Figure \ref{F:asymptotic} for the true concordance parameters $r_0^{(0)}=r_1^{(0)}= 0.925$. The other concordance parameter values exhibited similar trends.
 
In the left panel of Figure \ref{F:asymptotic}, the boxplots display the estimated percentages of correctly clustered covariate pairs, $100\hat{\tau}$, for the 25 simulated datasets.   We find that the accuracy of the inferred covariate-cluster allocations  increases with data dimension. The right panel of Figure \ref{F:asymptotic} demonstrates the methodology's success in discriminating between PDPs and Dirichlet processes;   the log-Bayes factor lower bounds increasingly  favor the true PDP model.

\subsubsection{Data generated by a different mechanism}\label{S:simulation_different}

We  evaluated  inference accuracy under model misspecification.
Twenty five  datasets were generated using a different process than  BaCon. Then, each dataset was  analyzed using the  BaCon methodology and the inferred clusters were compared with the true clusters.  Specifically,
for a true correlation  $\varphi^{(0)} \in\{0.90, 0.95, 0.99\}$ determining   within-cluster tightness, the  binary values of $n=100$ individuals  and $p=250$ covariates were generated as follows:

\begin{enumerate}

 \item \textbf{True  number of clusters:} Integer $Q_0$ was  generated from a Poisson distribution with mean $p/8$ and restricted to integers less than $\lfloor p/4\rfloor$.

 \item \textbf{True allocation variables:} 
 From among the distinct partitions of $p$ objects  consisting of exactly $Q_0$ sets, a partition was uniformly and randomly generated using the R package, {\tt  rpartitions}. Allocations variables
 $c_1^{(0)},\ldots,c_p^{(0)}$  compatible with this partition were randomly generated.    

\item \textbf{Multivariate normal vectors:} Given the cluster allocation variables, for individuals $i=1,\ldots,n$, we independently generated  row vectors $\boldsymbol{z}_i^{(0)}$ of length $p$ from a multivariate normal distribution with mean zero and $p \times p$ variance matrix $\boldsymbol{\Lambda}=((\Lambda_{j_1j_2}))$, where element
\begin{equation}
    \Lambda_{j_1j_2}=\varphi^{(0)}\, \mathcal{I}\bigl(c_{j_1}^{(0)}=c_{j_2}^{(0)}\bigr)+(1-\varphi^{(0)})\, \mathcal{I}\bigl(j_1=j_2\bigr). \label{Lambda}
\end{equation} The construct implies that $\text{Var}(z_{ij}^{(0)})=1$, so that $\text{Corr}\bigl(z_{ij_1}^{(0)}, z_{ij_2}^{(0)}\bigr)=\Lambda_{j_1j_2}$. In particular, the within-cluster correlations of the $z_{ij}^{(0)}$ 
are equal to $\varphi^{(0)}$, whereas   $z_{ij}^{(0)}$ belonging to different clusters are  independent.

\item \textbf{Binary covariates:} For individual $i=1,\ldots,n$ and  $j=1,\ldots,p$, we set the binary covariate 
$x_{ij}  = \mathcal{I}\bigl(z_{ij}^{(0)}  > 0\bigr)$.
\end{enumerate}

  Each artificial dataset was analyzed using the BaCon and k-means methodologies. The results were relatively robust to the correlation parameter $\varphi^{(0)}$.  
   Averaging over the 25 independent replications, Table \ref{T:varkappa5} displays the estimated percentage of correctly clustered covariate pairs, $\hat{\tau}$, for different values of correlation parameter.  In every  dataset, the  number of clusters  detected by BaCon was exactly equal to the true number of clusters, $Q_0$. Despite the considerably different data generation mechanism, we found that BaCon was highly accurate in detecting the underlying cluster structure and significantly outperformed the k-means algorithm.
   
    For the 
     lower bound of the log-Bayes factor,  previously introduced in Section \ref{S:simulation_PDP}, the second column of  Table \ref{T:varkappa10} displays  averages and standard deviations of the MCMC estimates for the 25 datasets.
In every situation, the  lower bounds of the Bayes factors are significantly greater than $e^6=403.4$, implying that the data overwhelmingly  favor  PDP~allocations over Dirichlet processes.
  Column 3 of  Table \ref{T:varkappa10} displays the 95\% posterior credible intervals for the PDP discount parameter, $d$, revealing that no posterior probability was assigned to Dirichlet process~models.

\begin{table}
\caption{When the data were generated using the procedure described in Section \ref{S:simulation_different},  the proportion of correctly clustered covariate pairs for two competing methods and different values of the true correlation parameter $\varphi^{(0)}$ in equation (\ref{Lambda}). The standard errors are shown in  parentheses.} \label{T:varkappa5}
\vspace{.2 in}
\centering
\begin{tabular}{ c   | c |c  }
\hline\hline
 \textbf{Correlation} &\multicolumn{2}{c}{\textbf{Percent $\hat{\tau}$}} \\
  \cline{2-3}
 \textbf{$\varphi^{(0)}$} &\textbf{BaCon}  &\textbf{K-Means}\\
\hline
0.90   &98.093 (0.329) &94.425  (0.477) \\
0.95 &98.448 (0.253) &94.702 (0.459) \\
0.99 &97.982 (0.315) &94.802 (0.349) \\
\hline\hline
\end{tabular}
\end{table}

\begin{table}
\caption{When the data were generated using the  Section \ref{S:simulation_different} procedure,   column 2 presents the average lower bound of the log-Bayes factor of PDP models, relative to Dirichlet process models, for different values of the true correlation parameter $\varphi^{(0)}$ in equation (\ref{Lambda}). Standard deviations for the 25 independent replications are shown in  parentheses. Column 3 displays  95 \%  posterior credible intervals for the PDP discount parameter $d$. }
\label{T:varkappa10}
\vspace{.2 in}
\centering
\begin{tabular}{ c   | c  | c}
\hline\hline
 \textbf{Correlation}  &\textbf{Lower  bound } &\textbf{95\% C.I.}\\
 \textbf{$\varphi^{(0)}$}  &\textbf{of  log-BF} &\textbf{for $d$} \\
\hline
0.90  &18.598  (6.930) &(0.178, 0.403)\\
0.95  &16.081 (4.595) &(0.164, 0.392)\\
0.99 &13.949 (4.714) &(0.149, 0.373)\\
\hline\hline
\end{tabular}
\end{table}

\begin{figure}
\begin{center}
\includegraphics[scale=0.41]{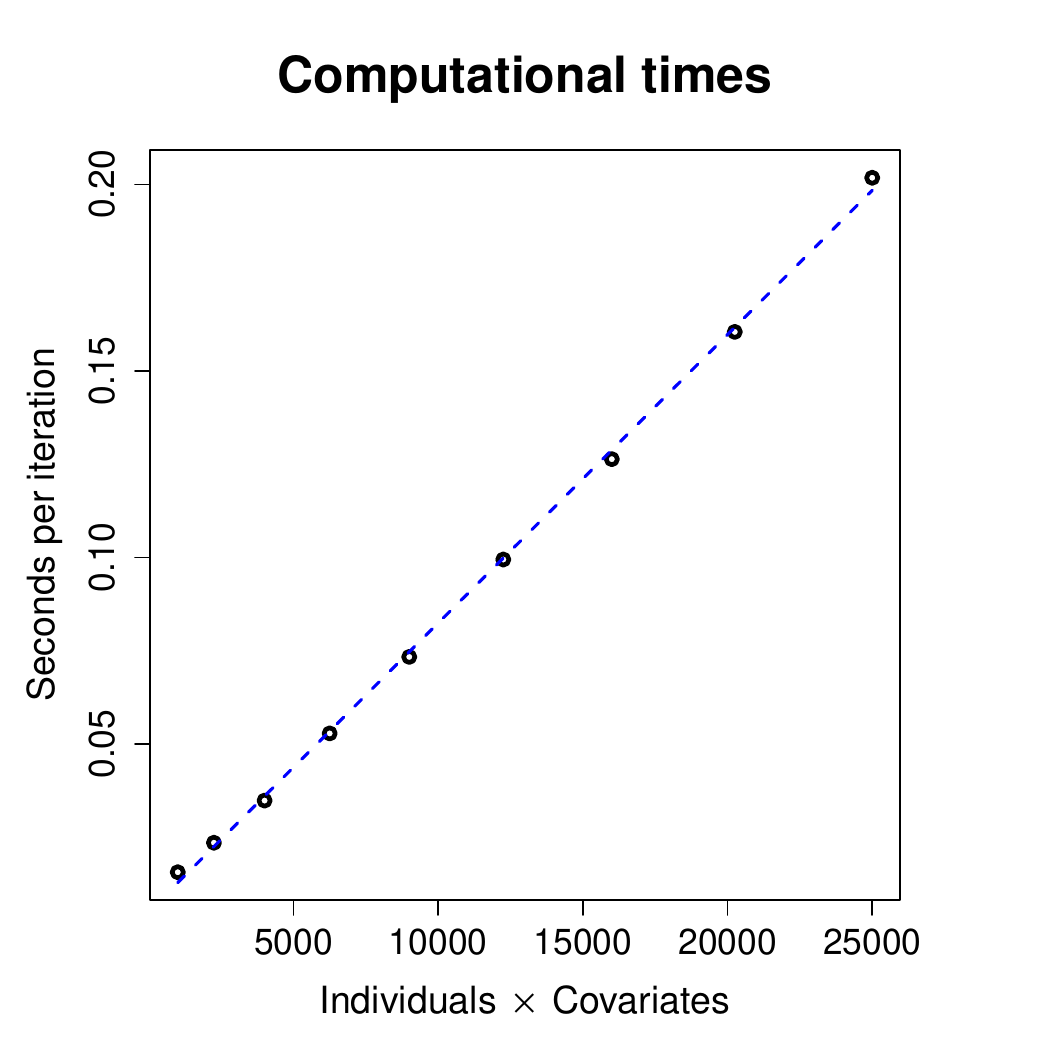}
\caption{Computational times for artificial datasets of increasing dimension generated using the process of Section \ref{S:simulation_different} with true correlation parameter $\varphi^{(0)}=0.9$. See the discussion in Section \ref{S:simulation_different}.}
\label{F:asymptotic2}
\end{center}
\end{figure}

 \paragraph{Computational complexity}\quad  For data generated using a true correlation  of $\varphi^{(0)}=0.9$, we compared the MCMC computational times as  the number of individuals  $n$ and covariates $p$ increased. All calculations were performed using the University of Florida HiPerGator2 supercomputer, which  has 30,000 cores in Intel E5-2698v3 processors with 4 GB of RAM per core and a total storage of 2 petabytes.  
  For the  $(n,p)$ pairs,   $(20,50)$, $(30,75)$, $(40,100)$, $(50,125)$, $(60,150)$, $(70,175)$, $(80,200)$, $(90,225)$,  and $(100,250)$,  Figure \ref{F:asymptotic2} plots the average computational time per MCMC iteration versus $np$, the total number of  matrix $\boldsymbol{X}$ elements. The best-fitting straight line is displayed for comparison. For  $V$ brain regions, the plot  suggests a computational cost of $O(np)$ or $O(n V^2)$ for the MCMC algorithm of Section \ref{S:MCMC}. True   correlations of 0.95 and 0.99 had nearly identical results.

\subsection{Prediction accuracy}\label{S:simulation1}

We assessed the prediction accuracy of our methods using $n=114$ artificially generated continuous responses. However, unlike the previous simulation studies,  we used the actual $p=1,374$ covariates from the  MRN-114 dataset. The following procedure was used to  generate and analyze 25 sets of subject-specific responses:

\begin{enumerate}

\item Randomly select $10$ covariates with mutual taxicab distances lying between 0.4 and 0.6. This gives the true predictor set   $\mathcal{S} \subset\{1,\ldots,p\}$   consisting of $|\mathcal{S}|=10$ members. Recall that for binary covariates, taxicab distances near 0 and 1  correspond  to high positive and negative correlations, respectively. Restricting these distances to a neighborhood of 0.5  avoids collinearity in the true predictors, which are unknown at the analysis stage. Nevertheless, there is high collinearity among the $p=1,374$ \textit{potential} predictors.

\item For each $\beta^*\in \{0.5, 0.85, 1.2\}$:

\begin{enumerate}

    \item[] For every individual indexed by $i=1,\ldots,n$, generate Gaussian responses $y_i$ with mean~$\beta^*/2 + \beta^*  \sum_{j\in \mathcal{S}}  x_{ij}$ and  standard deviation $\sigma_0=0.5$. The signal-to-noise ratio in the data increases as $\beta^*$ increases, with higher values of $\beta^*$ corresponding to higher associations between the response and true predictors.

\end{enumerate}

\item Randomly assign 91 individuals (roughly 80\%)  to the training set and the remaining individuals to the test set.

\item  Apply the  BaCon procedure for Gaussian responses to analyze the data. Choose a representative from each cluster  as described in option \textit{(a)} of Section \ref{S:predictors}. Make posterior inferences using the training data and predict the responses for the test~case individuals.

\end{enumerate}

We fitted the
 same simulated datasets using the techniques, Lasso \citep{Tibshirani_1997}, $L_2$-boosting \citep{Hothorn_Buhlmann_2006}, elastic net \citep{Zou_Trevor_2005}, and random forests \citep{Breiman_2001}.   These machine learning techniques are extensively used for binary predictor selection of continuous responses and have been implemented in the R packages glmnet, mboost, and randomForest. We used the default recommended settings for the tuning parameters of these R packages. In choosing these methods, we  focused on techniques capable of delivering  sparse, interpretable models and quantifying the effects of important brain region pairs. 

Because the artificial responses are continuous, the  prediction errors of the competing methods were compared using their percentage MSE reduction relative to the null model in the $n^*=13$ test case individuals. For a given dataset and method, the percentage MSE reduction is equal to $1-\sum_{i=1}^{n^*} (y_i - \hat{y}_i)^2/\sum_{i=1}^{n^*} (y_i - \bar{y})^2$,
where $\hat{y}_i$ is the method's predicted response for individual $i$.  A large reduction is indicative of a method's high prediction accuracy.

As a straightforward and transparent competitor to the proposed technique, we  applied the k-means algorithm to  group the $p$ columns of the matrix $\boldsymbol{X}$ into fewer, say $q^*$, number of concordant clusters, with $q^*$ chosen to maximize the median percentage MSE reduction over the range $q^* \le p/4$.  Next, for each k-means cluster, we computed the \textit{median potential predictor} as the covariate having the smallest sum of distances to the remaining covariates belonging to the cluster. Finally, from this smaller set of potential predictors, the set of predictors, along with their relationship with the responses, were inferred via $L_2$-boosting. We simply refer to this technique as ``K-means''.

\begin{table}
\caption{For  different~$\beta^*$, a comparison of the true positive rate (TPR) and true negative rate (TNR) of the  BaCon and K-Means methods.}\label{T:tpr}
\centering
\begin{tabular}{ c   | c |c  | c |c}
\hline\hline
 \textbf{Coefficient} &\multicolumn{2}{c}{\textbf{TPR}} &\multicolumn{2}{c}{\textbf{TNR}} \\
  \cline{2-5}
 \textbf{$\beta^*$} &\textbf{BaCon}  &\textbf{K-Means}&\textbf{BaCon}  &\textbf{K-Means}\\
\hline
0.50 &54.970 (3.553) &5.600  (1.536) &97.268 (0.279) &93.006 (0.218)\\
0.85 &88.536 (3.195) &7.600 (1.759) &99.189 (0.263)&93.473 (0.199)\\
1.20 &94.208 (1.276) &8.000 (1.732) &99.657 (0.132)&93.210 (0.262)\\
\hline\hline
\end{tabular}
\vspace{3mm}
\end{table}

Table \ref{T:tpr} displays the true positive and negative rates for the procedures BaCon and K-Means. For each method, the rates are computed under the notion that we are unable to distinguish between predictors assigned to the same cluster by that method. We find that, for all three levels of the association parameter $\beta^*$, the procedure BaCon  provides far more accurate inferences than the  K-Means~procedure.

\begin{figure}
\centering
\includegraphics[scale=0.31]{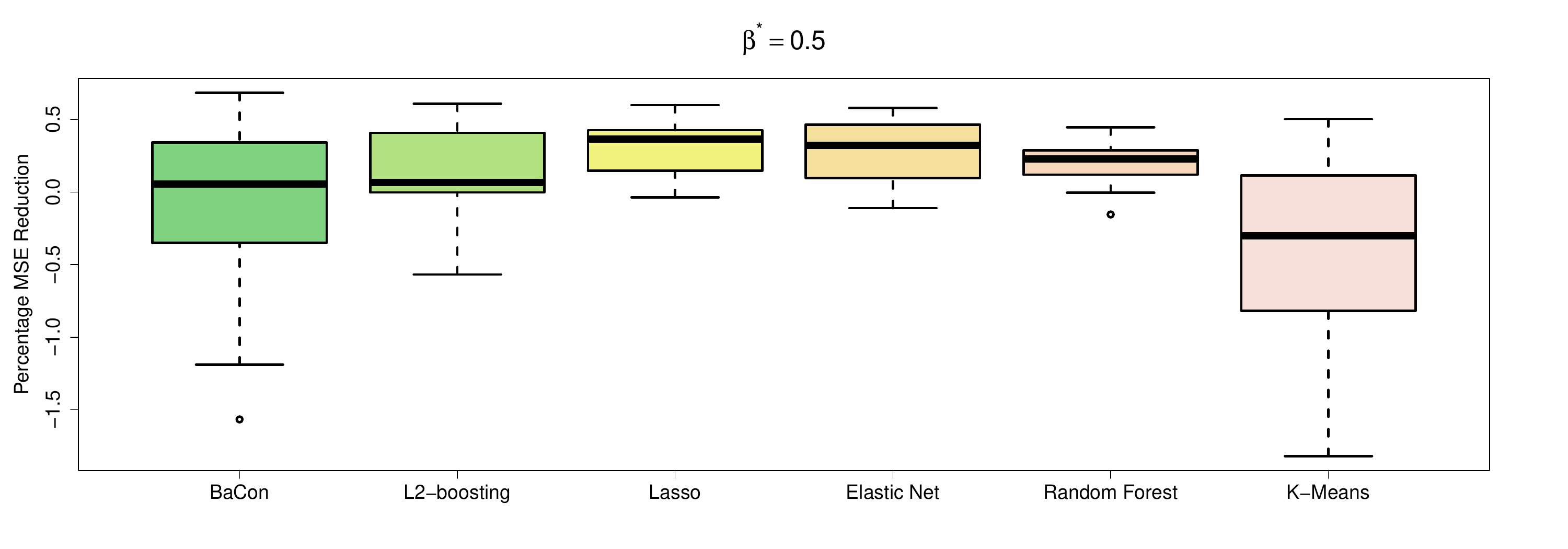}
\includegraphics[scale=0.31]{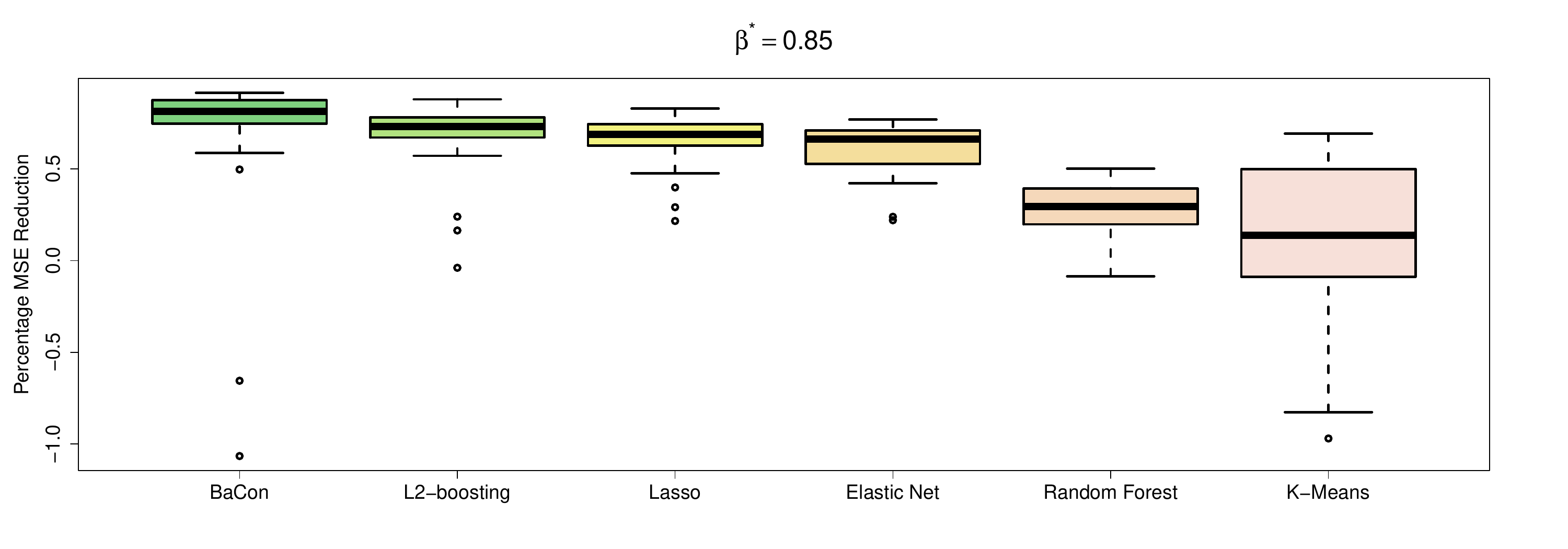}
\includegraphics[scale=0.31]{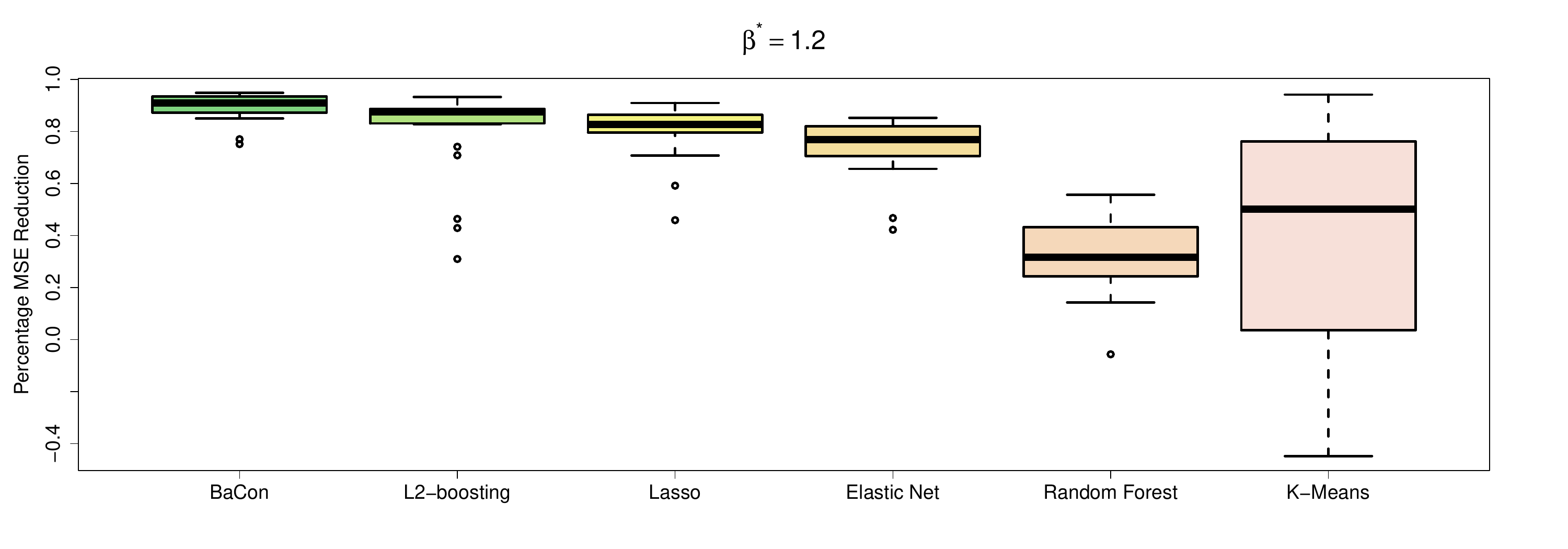}
\caption{Side-by-side boxplots comparing the prediction accuracy of the competing techniques in the simulation study of Section \ref{S:simulation1}.}\label{F:simulation2}
\end{figure}

\begin{table}
\caption{Comparison of the detected model sizes in the simulation study for different~$\beta^*$.}\label{T:simulation2}
\centering
\begin{tabular}{ l   | c |c  |c}
\hline\hline
 &$\beta^*=0.5$&$\beta^*=0.85$&$\beta^*=1.2$\\
\hline
\textbf{BaCon} &11.66 &10.32 &10.23\\
\textbf{$L_2$-boosting} &36 &30 &27\\
\textbf{Lasso} &54 &62 &71\\
\textbf{Elastic net} &93 &94 &95\\
\textbf{K-Means} &23 &24 &25\\
\hline\hline
\end{tabular}
\end{table}

Figure~\ref{F:simulation2} depicts  boxplots of the percentage MSE reductions for the different methods. As expected,
  the median percentage MSE reductions decrease for most procedures as $\beta^*$ increases. The only exception is random forests, for which the MSE reductions  essentially remain unchanged. The  K-means procedure has the highest variability. Irrespective of $\beta^*$, the K-means procedure often has high negative percentage MSE reductions,  rendering it unusable in  practice.

  When the true association between the response and true predictors is the weakest (i.e., when $\beta^*=.5$), Lasso performs the best. On the other hand, when the true association is non-negligible, BaCon is the clear winner.
 Table \ref{T:simulation2} displays the median number of predictors for each method. Being an MCMC sample average, the estimated model size for BaCon is typically a non-integer. We found that BaCon selects the sparsest models by far, and  irrespective of $\beta^*$, the detected model size approximately matches the true model size. The  other methods detected significantly overfitted models.

 In summary, we find that for most reasonable levels of predictor-response association, BaCon strikes the best balance between balances sparsity and prediction. It outperforms competing techniques,  with its gains dramatically increasing with the degree of predictor-response association.

\section{Data Analysis}\label{S:example}


Next, we analyzed the motivating  MRN-114  connectome dataset. We performed
25 independent replications of the following  steps: \textit{(i)} The data were randomly split in a 4:1 ratio into training and test sets.  \textit{(ii)} For the training cases, we analyzed the relationship between the CCI responses  and  pairwise brain region connectivity  as potential predictors using the techniques BaCon, $L_2$-boosting, Lasso, elastic net, and random forests. \textit{(iii)}~The five techniques were used to predict the  CCI responses of the test cases. For the BaCon procedure,  a single covariate representative from each cluster was selected in every MCMC iteration, as  described in option \textit{(a)} of Section \ref{S:predictors}. 

\begin{figure}
\begin{center}
\includegraphics[scale=0.31]{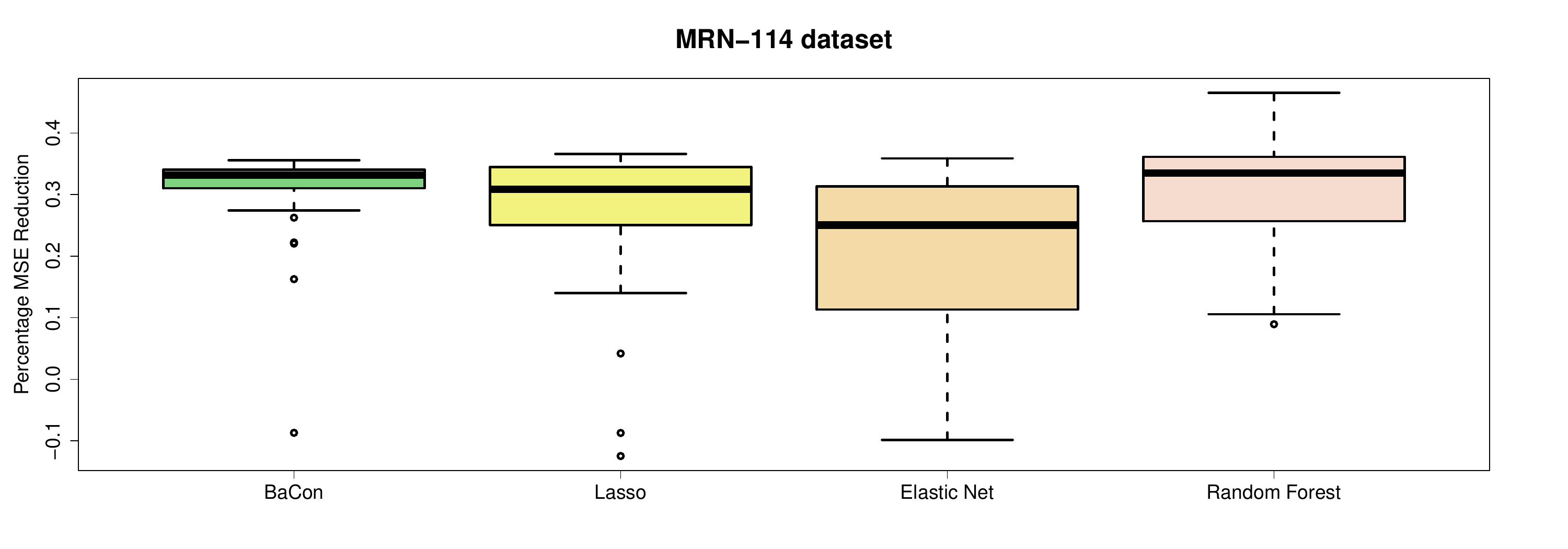}
\caption{Side-by-side boxplots comparing the prediction accuracies of different techniques.}
\label{F:simulation4}
\end{center}
\end{figure}

Figure \ref{F:simulation4} displays side-by-side boxplots of the percentage MSE reductions for the different methods. The accuracy and reliability of BaCon are significantly greater than those of Lasso and elastic net. The  random forests technique has the highest median accuracy, although it   displays fairly high volatility. The results for $L_2$-boosting are not shown in the figure because it had a significantly worse performance and a negative median MSE reduction.

\begin{figure}
\begin{center}
\includegraphics[scale=0.31]{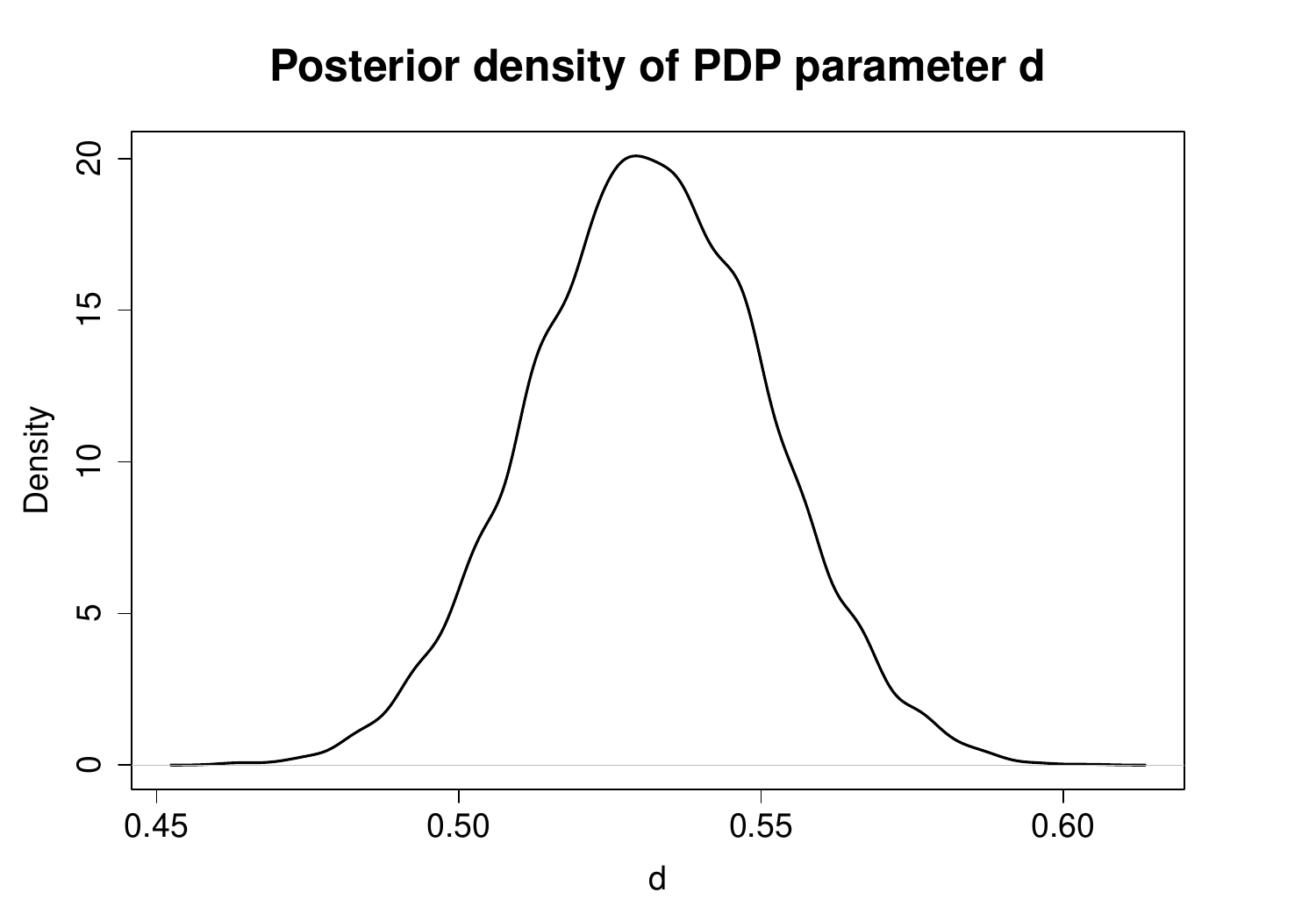}
\caption{Posterior summaries for the MRN-114 dataset.
}
\label{F:clustering}
\end{center}
\end{figure}

\begin{figure}
\begin{center}
\includegraphics[scale=0.3]{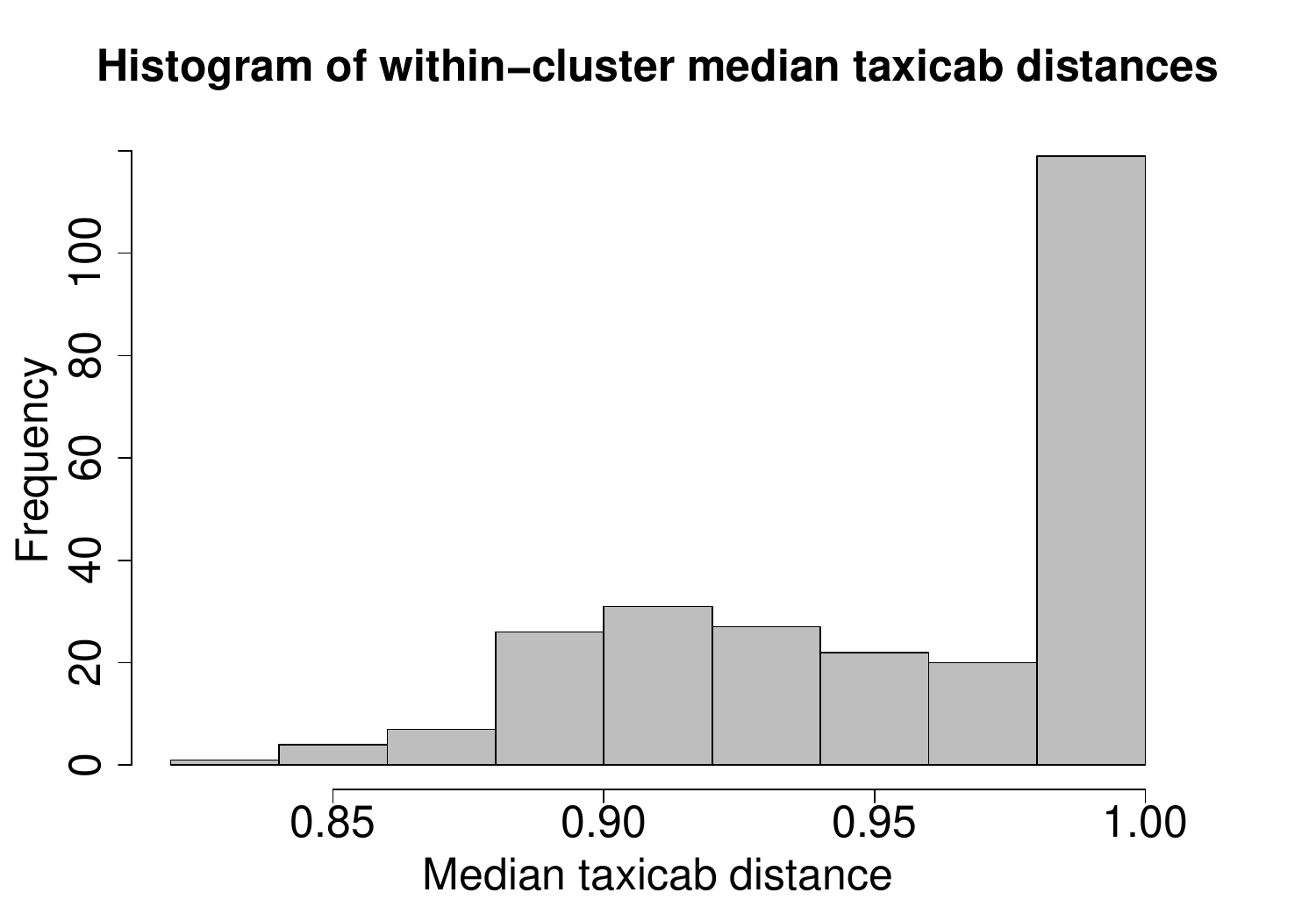}
\includegraphics[scale=0.3]{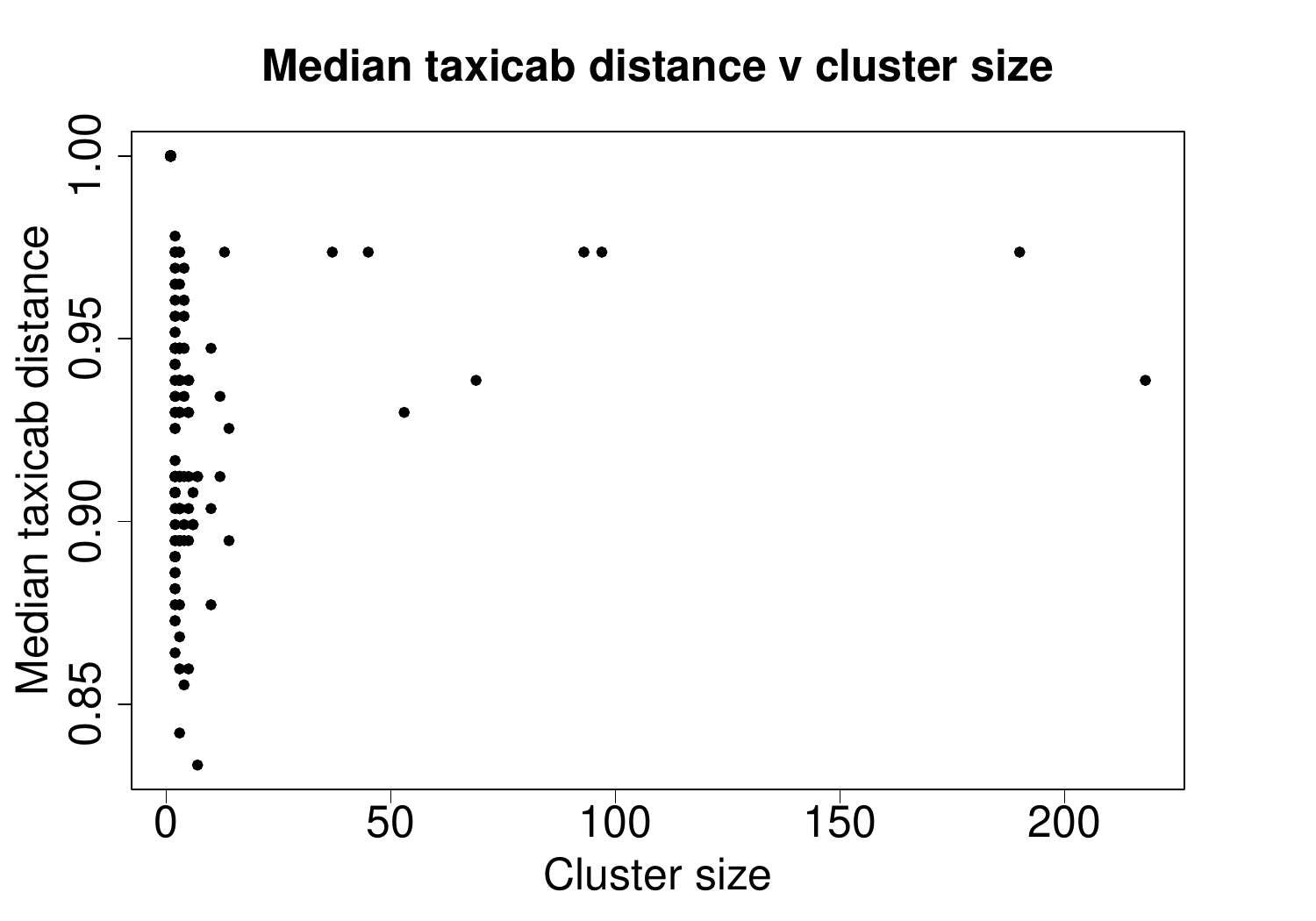}
\caption{For the MRN-114 dataset, median taxicab for the $\hat{q}=257$ PDP clusters of the least-squares allocation.}
\label{F:taxicab}
\end{center}
\end{figure}

The estimated marginal posterior density of the  PDP discount parameter $d$ is displayed in Figure \ref{F:clustering}. The  posterior probability of the event $[d=0]$ is estimated to be exactly zero. This suggests that a non-Dirichlet PDP allocation is strongly supported, as previously suggested  by the EDA.
As mentioned in Section \ref{S:MCMC}, we computed
the \textit{least-squares  allocation} for the covariate-to-cluster assignments.
The number of clusters  in the least-squares  allocation was $\hat{q}=257$.
For each least-squares allocation cluster, we computed the taxicab distances between the member covariates and the latent vector. The cluster-specific median distances are plotted in Figure \ref{F:taxicab}. The plots reveal high within-cluster concordance irrespective of cluster size, with the largest clusters having a higher-than-average median taxicab distance. These results demonstrate the effectiveness of BaCon as a model-based clustering procedure.

\begin{table}
\caption{Top seven clusters of pairs of brain regions that are most predictive of CCI. Each brain region pair in a cluster is listed along with its posterior probability of being a cluster representative. See the text for further discussion.}\label{T:dataanalysis}
\centering
	\label{table: LPML}
	\begin{tabular}{l | ll | c}
	\toprule
	\textbf{Cluster} & \textbf{Region 1} &  \textbf{Region 2} & \textbf{Probability} \\
	\midrule
	1 &lh-parsopercularis &lh-entorhinal     &0.978\\
1  &lh-parsopercularis  &rh-parsorbitalis   &0.010\\
1  &lh-parsopercularis  &rh-entorhinal   &0.006\\
1    &lh-parsopercularis  &rh-rostralanteriorcingulate   &0.003\\
1  &lh-inferiorparietal  &rh-bankssts   &0.003\\
\hline
2    &lh-parsorbitalis  &lh-superiorparietal  &1.000\\
\hline
3   &lh-caudalmiddlefrontal  &lh-lateralorbitofrontal  &1.000\\
\hline
4   &rh-parsopercularis  &rh-temporalpole   &0.941\\
4    &rh-parsopercularis  &rh-precentral   &0.021\\
4    &rh-parsopercularis  &rh-supramarginal   &0.019\\
4   &rh-parsopercularis  &rh-isthmuscingulate    &0.019\\
\hline
5  &lh-middletemporal  &lh-paracentral  &1.000\\
\hline
6    &lh-rostralmiddlefrontal  &lh-MeanThickness  &1.000\\
\hline
7    &lh-medialorbitofrontal  &rh-precuneus   &0.764\\
7    &lh-medialorbitofrontal  &rh-superiortemporal   &0.072\\
7    &lh-superiorfrontal  &rh-precuneus   &0.038\\
7    &lh-superiorfrontal  &rh-caudalmiddlefrontal   &0.035\\
7    &lh-medialorbitofrontal  &rh-caudalmiddlefrontal   &0.033\\
7    &lh-superiorfrontal  &rh-cuneus   &0.029\\
7    &lh-superiorfrontal  &rh-superiortemporal   &0.029\\
	\bottomrule
	\end{tabular}
\end{table}

Table \ref{T:dataanalysis} lists seven clusters of pairs of brain regions according to the Desikan atlas that are most predictive of composite creativity index (CCI), for which the cluster-level posterior probabilities of being predictors exceeded 0.8. Although the cluster labels  are arbitrary,  the clusters appear in decreasing order of  posterior probability of being (cluster) predictors, e.g.,  cluster 1 is the most predictive of CCI.
Each cluster  consists of one or more brain region pairs, e.g., cluster 1 consists of 5  region pairs, whereas clusters 2 and 3 consist of one pair each.
Within each cluster, each brain region pair (i.e., covariate) is listed along with its posterior probability of being a cluster representative. For example, the most important brain region pair in cluster 1 is the left hemisphere pair consisting of the regions
lh-entorhinal and lh-parsopercularis with a cluster representative posterior probability of 0.978.

These results confirm the findings of \citet[Table 1]{Jung_etal_2010}, who  detected regions within the lingual, cuneus, inferior parietal, and
cingulate brain regions corresponding to Table \ref{T:dataanalysis}. Specifically, regions within the so-called ``default mode'' network  are generally associated with creative cognition;
particularly, divergent thinking associated with CCI (e.g., medial frontal, precuneus). Our findings are also consistent with the review paper by \cite{10.3389/fnhum.2013.00330} that first outlined structural regions comprising the default mode network underlying creative cognition.
Finally, a recent meta-analysis \citep[Tables V and VI]{Wu_etal_2015} showed both structural and functional correlates of DTT
(divergent thinking tasks like CCI) which overlap significantly with our findings.

While the current approach largely supported previous research linking creative cognition to structure and function within the default mode network, other regions were elucidated by this methodology that have not been previously described within structural neuroimaging studies of creative cognition; see \cite{10.3389/fnhum.2013.00330} for a review. For example, the preponderance and strength of findings within bilateral inferior frontal lobe, particularly pars opercularis, are relatively novel within the creativity neurosciences. One study of patients suffering from lesions to various brain regions found that lesions to the left inferior frontal gyrus (including pars opercularis and pars triangularis), were found to exhibit high originality scores on divergent thinking tasks \citep{Shamay-Tsoory-etal}, suggesting that this hub might be critical to modulation of creative generation. Given that the left inferior gyrus is critical to processing verbal information \citep{Gernsbacher-etal}, this region is also likely to be critical to performance across tasks that are dependent upon verbal output, upon which the vast majority of divergent thinking tasks depend. Support for this notion is found in a study that found regional gray matter volume within the left inferior frontal gyrus (BA 45 - pars opercularis) to be associated with verbal creativity on a divergent thinking task \citep{Zhu_Zhang_Qiu}. 

Given that pars opercularis was most often paired with other brain regions in predicting CCI, the BaCon methodology has revealed a central ``hub'' from which creative cognition (particularly, modulation of originality) might derive. This potential hub has not been previously described in the creativity neuroscience literature, and warrants further research.

\section{Conclusions}\label{S:conclusion}

We focus on the problem of developing accurate predictive models for cognitive traits and neuro-psychiatric disorders using  an individual's brain connection network.
We have introduced a class of Bayesian Connectomics (BaCon) models that  rely on Poisson-Dirichlet processes to detect a lower-dimensional, bidirectional (covariate, subject)
 pattern in the adjacency matrix defining  brain region connections. This facilitates  effective stochastic searches, improved inferences, and test case predictions via  spike-and-slab priors for the lower-dimensional  cluster predictors. In simulation studies and analyses of the motivating connectome dataset, we find that BaCon performs reliably and accurately compared to established statistical and machine learning procedures. The substantially higher  accuracy of BaCon more than compensates for  possibly higher clock-times relative to some  competitors.
 The data analysis  confirms  findings in the literature  that have detected associations between creative cognition and the lingual, cuneus, inferior parietal, and
cingulate brain regions. Additionally, 
the BaCon methodology  detects a previously unknown  focal point from which modulation of originality, and creative cognition in general, might possibly emanate.

The BaCon methodology in its current form assumes no missing connectivity information.  This is, in general, a reasonable assumption given current connectome reconstruction algorithms. A bigger problem than missing data is measurement errors and outlying connectomes due to head movement, problems we will address in future research. Due to the intensive MCMC computations, we have performed
the clustering and variable selection parts of BaCon in separate stages, with the second stage relying on the least-squares estimate of the clustering pattern obtained from the first stage.  However, the inference procedure is potentially scalable and can be implemented on massively parallel devices such as graphical processing units (GPUs) using fast MCMC algorithms.
This would facilitate fully Bayesian posterior inferences via scalable, single-stage implementations of~BaCon.  In the near future, user-friendly code will be made available on a GitHub repository and
through the OpenConnectome project.

\newpage

 \section*{Acknowledgments}\label{A:MCMC_Q}
This work was partially supported by the National Science Foundation under Award DMS-1854003 to SG, and by  grant R01MH118927 of the National Institute of Mental Health and grant R01ES027498 of the National Institutes of Environmental Health Sciences, both part of the United States National Institutes of Health, to DD.

 \section*{Appendix: Derivation of the Gibbs sampler for matrix~$\boldsymbol{Q}$}\label{A:MCMC_Q}

\paragraph{Updating concordance parameter vector $\boldsymbol{r}$.}
From equation (\ref{X}), we find that the conditional likelihood function of matrix~$\boldsymbol{Q}$ is
\begin{align}
\bigl[\boldsymbol{X} \mid \boldsymbol{Q}^*, \boldsymbol{r}, \cdots \bigr] &= \prod_{i=1}^n \prod_{j=1}^p q_{v_{ic_j} x_{ij}} = \prod_{s=0,1} \prod_{t=0,1} q_{st}^{n_{st}} \notag\\
&= \prod_{s=0,1} \left\{ (1-r_s)^{n_{s, 1-s}}\cdot (q_{s, 1-s}^*)^{n_{s, 1-s}}\right\} \times
        \prod_{s=0,1} \bigl(r_s + (1-r_s)q_{ss}^*\bigr)^{n_{ss}} \label{[X_temp]}
\end{align}
Applying the binomial theorem, we obtain
\begin{align*}
\bigl(r_s + (1-r_s)q_{ss}^*\bigr)^{n_{ss}} &= \sum_{v_s=0}^{n_{ss}} {n_{ss} \choose v_s} r_s^{v_s} (1-r_s)^{n_{ss}-v_s} (q_{ss}^*)^{n_{ss}-v_s}\\
&= (1-r_s)^{n_{ss}} \sum_{v_s=0}^{n_{ss}} {n_{ss} \choose v_s} \rho_s^{v_s} (q_{ss}^*)^{n_{ss}-v_s}
    \quad \text{where $\rho_s=\frac{r_s}{1-r_s}$.}
\end{align*}
Substituting  into equation (\ref{[X_temp]}) above gives
\begin{equation}
\bigl[\boldsymbol{X} \mid \boldsymbol{Q}^*, \boldsymbol{r}, \cdots \bigr]
= \prod_{s=0,1} \left\{ (1-r_s)^{N_s} \sum_{v_s=0}^{n_{ss}} {n_{ss} \choose v_s} \rho_s^{v_s} (q_{ss}^*)^{n_{ss}-v_s}(q_{s,1-s}^*)^{n_{s,1-s}} \right\}. \label{[X|Q*,r]}
\end{equation}
Now the prior for $\boldsymbol{Q}^*$ in expression (\ref{Q_rows}) is
\begin{equation}
[\boldsymbol{Q}^*] = \prod_{s=0,1} \frac{1}{B(\frac{\alpha}{2}\boldsymbol{1})} \prod_{t=0,1} (q_{st}^*)^{\frac{\alpha}{2}-1} \label{[Q*]}
\end{equation}
Multiplying equations (\ref{[X|Q*,r]}) and (\ref{[Q*]}) and marginalizing over matrix $\boldsymbol{Q}^*$, we have
\begin{equation}
\bigl[\boldsymbol{X} \mid \boldsymbol{r}, \cdots \bigr]
= \prod_{s=0,1} \left\{  \sum_{v_s=0}^{n_{ss}} {n_{ss} \choose v_s} r_s^{v_s} (1-r_s)^{N_s-v_s} B(\boldsymbol{n}_s + \frac{\alpha}{2}\boldsymbol{1}-v_s\boldsymbol{1}_s)/B(\frac{\alpha}{2}\boldsymbol{1}) \right\}. \label{[X|r]}
\end{equation}

Let $f\left(\cdot \mid r_*, v_s+r_\alpha, N_s-v_s+r_\beta\right)$ be the density of the left-truncated beta distribution,  $ \text{beta}(v_s+r_\alpha,N_s-v_s+r_\beta)\cdot \mathcal{I}(r_*,\infty)$.
Multiplying the truncated beta priors for the concordance parameters in specification (\ref{Q_rows})  with likelihood expression (\ref{[X|r]}), and including  appropriate normalizing constants, we find that the full conditional of the concordance parameters $\boldsymbol{r}$ is
\begin{equation}
\bigl[\boldsymbol{r} \mid \boldsymbol{X}, \cdots \bigr]
= \prod_{s=0,1} \left\{  \sum_{v_s=0}^{n_{ss}} h_s(v_s) \cdot f\left(r_s \mid r_*, v_s+r_\alpha, N_s-v_s+r_\beta\right) \right\} \label{[r|X]2}
\end{equation}
for the pmf $h_s(\cdot)$ in definition (\ref{h_s}). This  is equivalent to  full conditional (\ref{[r|X]}).

\paragraph{Updating matrix $\boldsymbol{Q}^*$ conditional on concordance parameter vector $\boldsymbol{r}$.}
Assuming vector $\boldsymbol{r}$ to be known, we multiply equations (\ref{[X|Q*,r]}) and (\ref{[Q*]}) and normalize to obtain
\begin{equation}
\bigl[\boldsymbol{Q}^* \mid \boldsymbol{X}, \boldsymbol{r}\cdots \bigr]
= \prod_{s=0,1} \left\{  \sum_{v_s=0}^{n_{ss}} l_s(v_s) \cdot \partial_2\left(\boldsymbol{q}^*_s \mid \boldsymbol{n}_s + \frac{\alpha}{2}\boldsymbol{1}-v_s\boldsymbol{1}_s\right) \right\} \label{[Q*|r]}
\end{equation}
for the pmf $l_s(\cdot)$ of definition (\ref{l_s}),  with $\partial_2(\cdot | \boldsymbol{a})$ denoting the density of the Dirichlet distribution, $\mathcal{D}_2(\boldsymbol{a})$. This is equivalent to  full conditional (\ref{[Q*|X,r]}).


\newpage


\begin{spacing}{0.9}


\bibliographystyle{asa}
{
\small
\bibliography{main}
}


\end{spacing}

\end{document}